\documentclass[12pt,preprint]{aastex}
\usepackage{morefloats}
\usepackage{color}

\newcommand{\unit}[1]{\,{\rm #1 }}
\newcommand{\mtr}{\unit{m}}
\newcommand{\kpc}{\unit{kpc} }
\newcommand{\arcsecsq}{\unit{arcsec^2}}
\newcommand{\arcsecsqi}{\unit{arcsec^{-2}} }
\newcommand{\arcseci}{\unit{arcsec^{-1}} }
\newcommand{\hubbleunit}{\unit{km\,sec^{-1}\,Mpc^{-1}}}
\newcommand{\magarcsecsqi}{\unit{mag\arcsecsqi}}
\newcommand{\angstrom}{\unit{\rm{\AA}}}
\newcommand{\angstrompixi}{\unit{\rm{\AA\,pixel}^{-1}} }
\newcommand{\absm}[1]{M_{\rm#1}}
\newcommand{\appm}[1]{m_{\rm#1}}
\newcommand{\deri}[2]{\frac{d#1}{d#2}}

\newcommand{\bbs}{\left[}
\newcommand{\ebs}{\right]}
\newcommand{\bbb}{\Big(}
\newcommand{\ebb}{\Big)}
\newcommand{\bbbs}{\Big[}
\newcommand{\ebbs}{\Big]}

\newcommand{\bba}{\left<}
\newcommand{\eba}{\right>}

\newcommand{\mean}[1]{\bba #1 \eba}
\newcommand{\eg}{e.\,g.\,}

\newcommand{\labequn}[1]{\label{eq:#1}}
\newcommand{\labfig}[1]{\label{fig:#1}}
\newcommand{\labsecn}[1]{\label{sec:#1}}
\newcommand{\labsubsecn}[1]{\label{subsecn:#1}}

\newcommand{\labtablem}[1]{\label{tab:#1}}

\newcommand{\equn}[1]{Equation~\ref{eq:#1}}
\newcommand{\fig}[1]{Figure~\ref{fig:#1}}
\newcommand{\secn}[1]{Section~\ref{sec:#1}}
\newcommand{\subsecn}[1]{Section~\ref{subsecn:#1}}

\newcommand{\tablem}[1]{Table~\ref{tab:#1}}

\newcommand{\dequn}[2]{Equations~{\ref{eq:#1}}~and~{\ref{eq:#2}}}

\newcommand{\mfig}[2]{Figures~{\ref{fig:#1}}~to~{\ref{fig:#2}}}
\newcommand{\dtablem}[3]{Tables~\ref{tab:#1},~\ref{tab:#2}~and~\ref{tab:#3}}

\shorttitle{Isophotal Shapes}
\shortauthors{Chaware et al.}

\begin{document}


\title{Isophotal shapes of early-type galaxies to very faint
levels}


\author{Laxmikant Chaware,\altaffilmark{1}
Russell Cannon,\altaffilmark{2}
Ajit K.  Kembhavi,\altaffilmark{3}
Ashish Mahabal\altaffilmark{4}
\and 
S. K. Pandey\altaffilmark{1}
}
\altaffiltext{1}{School of Studies in Physics, Pt. Ravishankar Shukla University,
Raipur – 492010, India
}
\altaffiltext{2}{Australian Astronomical Observatory, PO Box 915,
North Ryde, NSW 1670, Australia.
}
\altaffiltext{3}{Inter University Center for Astronomy and Astrophysics, Pune, India
}
\altaffiltext{4}{California Institute of Technology,  
Pasadena, CA 91125, USA
}


\begin{abstract}
We report on a study of the isophotal shapes of early-type galaxies,  
to very faint levels reaching  $\sim0.1\%$ of the sky brightness. The 
galaxies are from the Large Format Camera (LFC) fields obtained using the
 Palomar $5\mtr$ Hale telescope, 
with integrated exposures ranging from 1 to 4 hours in the 
SDSS {\it r}, {\it i} and {\it z} bands. The shapes of isophotes of early-type 
galaxies are important as they are correlated 
with the physical properties of the galaxies and are influenced by galaxy 
formation processes. In this paper we report on a sample 
of 132 E and SO galaxies in one LFC field. We have redshifts for 53 of these, 
obtained using AAOmega on the Anglo-Australian Telescope. 
The shapes of early-type galaxies often vary with radius. 
We derive average values of isophotal shape parameters in 
four different radial bins along the semi-major axis in each 
galaxy. We obtain empirical fitting formulae for the probability 
distribution of the isophotal parameters in each bin and 
investigate for possible correlations with other global 
properties of the galaxies. Our main finding is that the 
isophotal shapes of the inner regions are statistically different 
from those in the outer regions. This suggests that the outer 
and inner parts of early-type galaxies have evolved somewhat independently. 
\end{abstract}




\section{Introduction}
The morphology and colours of galaxies carry important information 
about their formation and evolution. \citet{ben88} studied 
a sample of 109 large, nearby early-type galaxies
to examine their isophotal shapes, using Fourier expansions 
in the polar angle to distinguish between boxy (rectangular) isophotes and 
disky (pointed) isophotes.  Studies have shown that elliptical 
galaxies with disky isophotes tend to be fainter, rotationally 
supported, lack X-ray and radio activity and have power-law
nuclear light profiles,  while those with boxy isophotes tend to be brighter,
supported by random motions, have significant X-ray and radio activity
and peaked  nuclear profile \citep{fer94,bos94,res01,lau05}. The shapes of 
galaxies may also be correlated with their ages \citep{ryd01}. 
Recently \citet{hao06}  studied the isophotal shapes of a large sample of 
847 early-type galaxies from the SDSS,  and \citet{pas07}
have used the same sample to investigate the dependence of the
isophotal structure on the AGN activity and environment of the galaxies.
The dynamic range available to these studies is limited, due to the short
exposure of the SDSS.
The isophotal shape study by \citet{ben88} extends to 1.5 times effective 
radius (1.5$r_e$) of the galaxies, while \citet{hao06} studied their larger sample of 857 
nearby galaxies with redshift $z < 0.05$ extending to   
1.5 times the Petrosian half-light radius (1.5$R_{50}$).

In this paper we revisit the properties of isophotal shapes for a sample of 
more distant early-type galaxies in a 
single field, using exceptionally deep CCD exposures. These enable us to go out to
much larger radii of $\sim4.5R_{50}$ and reach $\sim4\magarcsecsqi$ deeper in 
surface brightness.  We have obtained redshifts for a subsample of these galaxies: 
the redshifts peak at $z\sim0.1$ and extend to $z \sim0.8$, covering a 
wide range of absolute magnitudes. 

Studies of the faint outer parts of galaxies are not new. By 1980 Malin \citep[e.g.][]{mal80} 
was using sky survey photographs to create images that reached 10 magnitudes below the 
level of the sky background. These led to the discovery of very faint shells and tails around
galaxies, soon interpreted as evidence for dynamical evolution due to 
gravitational interactions and mergers \citep[e.g.][]{qui84}. \citet{cap88} combined 
photographic and CCD imaging of nine nearby galaxies, to explore the boundary 
between E and S0 galaxies by looking at the changing shapes of isophotes with radius.

We extend the methodology of \citet{ben88} and derive mean isophote parameters 
for up to four radial bins for each galaxy.
Pointed isophotes can usually be attributed to a galaxy which consists of a
spheroid and a weak disk which is being viewed more or less edge-on.    
\citet{nie89} suggest that some pointed isophotes may be due to tidal
extensions. But the origin of boxy isophotes is still unclear. \citet{naa99}
showed that mergers of spiral galaxies with comparable mass are more likely to form 
boxy ellipticals, while mergers of spiral galaxies with differing mass can lead to 
disky ellipticals. However, disc-disc mergers do not form a perfect dichotomy
\citep{naa03}, while elliptical-elliptical mergers always lead to
boxy isophotes \citep{naa06}. Using N-body simulation \citet{bou05}
found that for a 7:1 merger, isophotes in the inner region (bulge) are  boxy, while 
the outer isophotes are disky. Such radial variation is also seen in the 
analysis of 2MASS data for  Arp mergers by \citet{chi02}.  Our study of
isophotal shapes to faint levels will be useful for better comparison between 
observations and such theoretical possibilities.  

This paper is organized in the following way:
in \secn{data} we describe the imaging data, sample selection, bulge-disk 
decomposition and spectroscopic data used for this study. 
In \secn{parameter estimation}, surface photometry of sample galaxies, 
and estimation of isophotal and rest frame parameters, are described.
In \secn{observed properties}, we give the properties of the sample and main results 
of our study along with the comparison with earlier work.
We discuss our results in \secn{discussion} and  
 give a  summary and our conclusions in  \secn{conclusion}.

\section{The observational data}
\labsecn{data}
\subsection{Imaging data}
\labsubsecn{imaging}
 
The imaging data 
were originally obtained as part of a faint quasar survey  using the Large 
Format Camera (LFC) on the $5\mtr$ Hale telescope at Mount Palomar 
Observatory,  with integrated  exposures ranging from 1 to 4 hours 
in {\it r}, {\it i} and {\it z} filters. The LFC is a mosaic of six $2048 \times 4096$ 
pixel CCDs, mounted at the prime focus of the Hale Telescope.
The fields targeted were those of known SDSS high redshift ($z>5$) quasars in 
order to look for a local over-density of quasars \citep{mah05}. 
We obtained  reduced images
which were bias subtracted, flat fielded and  processed to remove cosmic rays 
as a part of the original programme.
The deep LFC images have excellent signal-to-noise ratio (SNR),
permitting us to examine faint features, including the outermost 
extensions of the galaxies, down to $\sim0.1\%$ of the sky brightness.

In this paper we use a sample of  galaxies
selected from the LFC field SDSS 1208+0010 which spans an area of 
$13\arcmin.56\times25\arcmin.32$ centered on
$\rm{RA}=12^h08^m24^s$ and $\rm{dec}=00^d10^m28^s$ (J2000).
The field was observed in the SDSS {\it i} and {\it z} filters
with integrated exposure of $2.51\unit{h}$  and $2.56\unit{h}$ respectively. The 
{\it i}-band image of the field is an average combination of 140 individual frames.
The mean background count in the {\it i}-band image is 
6278 ADU (analog-to-digital units), so with a gain of 1.1, an
accuracy of $\sim0.1\%$ is obtained in the estimation of the sky background.   

All the LFC fields are in the area of the sky covered by 
SDSS imaging. We therefore have imaging data in the SDSS {\it u, g} and {\it r} bands
for our sample galaxies,
in addition to the LFC {\it i} and {\it z} band data, though not to the same depth.
We have used the SDSS DR7 photometric catalog \citep{aba09} to obtain 
the Petrosian half-light radii, apparent magnitudes and Galactic 
extinction corrections at the position of each sample galaxy in all
the five bands.

\subsection{Sample selection }
\labsubsecn{sample}

We used SExtractor \citep{ber96} to produce a catalog of all sources in the 
LFC field SDSS 1208+0010, and to identify galaxies from the catalog.   
SExtractor  provides a {\it stellarity index}  for
star/galaxy separation of the detected objects: 0 for an object
identified as a galaxy, 1 for a star, and some  intermediate 
value for  ambiguous objects, which are usually the fainter objects in
the sample.  For correct star/galaxy separation, and to have galaxies with
sufficient brightness and size to ensure reliable morphological study, we select
galaxies using objects from our SExtractor catalog which satisfy the following 
criteria:  (i) SExtractor parameter value $\rm{CLASS\_STAR} < 0.8$; (ii)
SDSS parameter value $\rm{PhotoType}=\rm{GALAXY}$ 
in the PhotoObjAll table of the SDSS DR7 photometric catalog
\citep{aba09}; and (iii) $\appm{i} < 20.5$ where $\appm{i}$
is the {\it i}-band model magnitude of the object corrected for extinction, which is 
denoted by {\it dered\_i} in the SDSS DR7 photometric catalog. To ensure that a 
sufficient number of pixels is available for surface photometry, we included in our sample
only objects with the {\it i}-band SExtractor parameter  ${\rm ISOAREA\_IMAGE}>165$.  The
area covered by the chosen pixels is $>21.74 \arcsecsq$ for every object in the sample,
with the intensity in every pixel being brighter than the analysis threshold of 
$26.0 \magarcsecsqi$, which corresponds to a $3 \sigma$ detection threshold. 
Through visual examination of the image, 
we have excluded objects which are near bright stars.  Applying these criteria, 
we arrived at a sample of 266 galaxies.

Basic data for the target galaxies in Field SDSS 1208+0010 are
available in \tablem{basic full}, which is at the end of this article.
We give first few entries from \tablem{basic full} in \tablem{basic} as  an illustration. 
The table gives an LFC serial number for
each galaxy, from the SExtractor catalog generated by us, the SDSS ID,
position and photometric data from from the DR7 photometric catalog
\citep{aba09}, and finally the semi-major axis (in arcsec), 
{\it i}-band surface brightness (in $\magarcsecsqi$) and
signal-to-noise ratio (SNR) of the outermost isophote to which an ellipse
can be fitted (see \secn{parameter estimation}). 

\begin{figure}
        \centering

\fbox{
\includegraphics[width=0.35\textwidth,angle=0.0,clip]{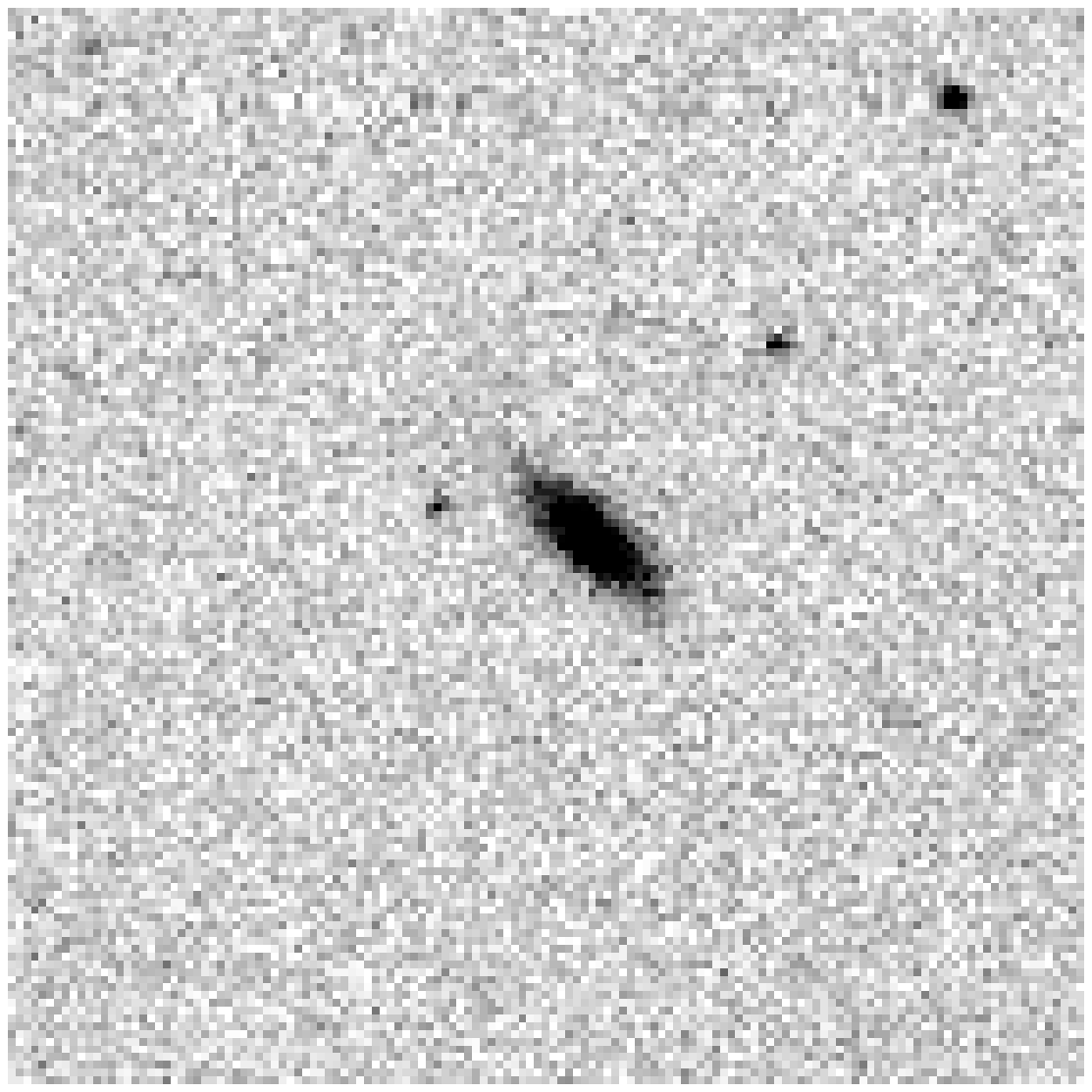}}%
\fbox{
\includegraphics[width=0.35\textwidth,angle=0.0,clip]{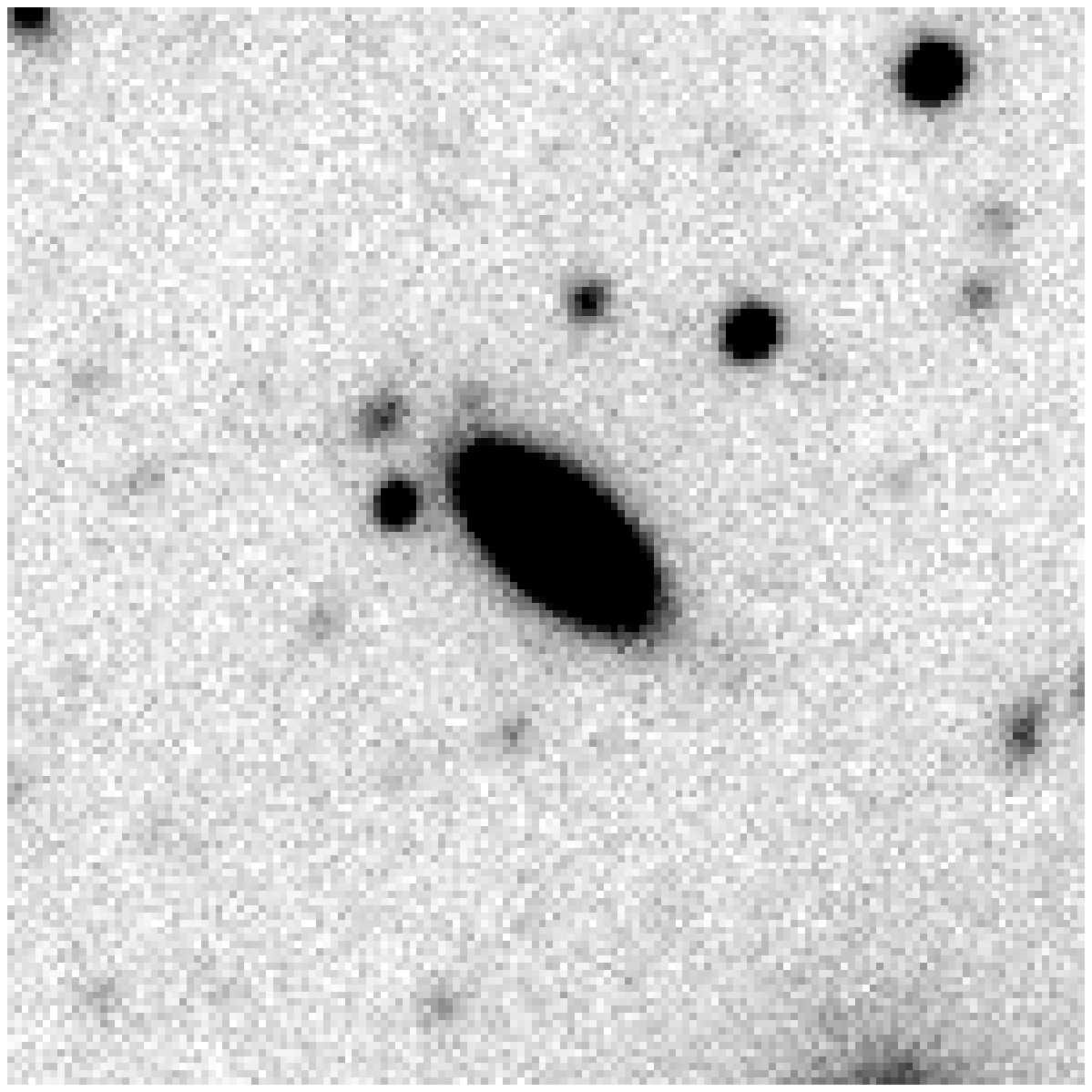}}

\includegraphics[width=0.7\textwidth,clip]{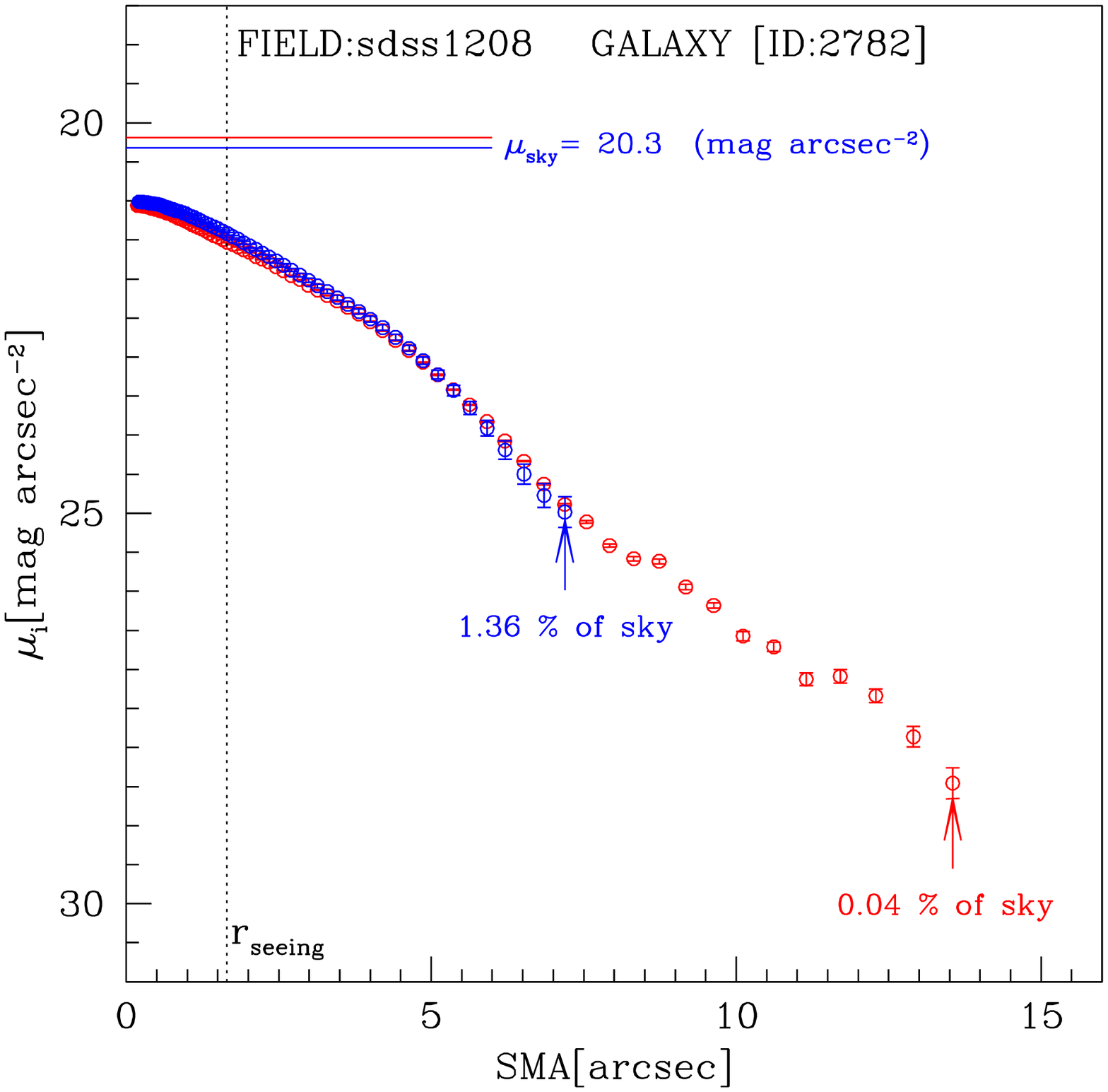}
\caption{ {\small Surface brightness (SB) profiles of 
a galaxy from the sample, generated using SDSS (shown in blue color) and
 LFC (shown in red color) {\it i} 
band images. The top panels show SDSS (left) and LFC (right)
image cutouts ($\sim 55 \arcsec \times 55 \arcsec$).
The SB profiles are obtained by fitting 
ellipses up to a point along the semi major axis where the ratio of the mean 
isophotal intensity to the total error in the isophotal intensity drops to three
in each image. $1 \sigma $ error in the isophotal intensity is indicated 
for each point. The dotted vertical line shows the average seeing in the {\it i} band 
LFC observation. The galaxy has $\appm{\it i}=17.8$, and $z= 0.1772$ as determined from
 AAT observation.  The linear scale for the cosmology adopted 
(see \secn{parameter estimation}) is $\sim3\kpc \arcseci$.} }
\labfig{compare}
\end{figure}

We show in \fig{compare} the {\it i}-band image of the largest galaxy from the 
LFC field, together with the SDSS {\it i}-band image of the same galaxy.
The profiles of the surface brightness along the semi-major axis of the two images are 
shown in the lower panels, up to a point where the ratio of the mean isophotal intensity
to the total error in isophotal intensity is about three. These
points are shown as arrows along the two profiles, with the surface brightness
level reached at that point given as a fraction of the sky brightness. Since
the SNR is the same for isophotes of the two images at the indicated points, 
it is clear that
we reach  $\sim4\magarcsecsqi$  fainter
in surface brightness in the LFC images compared to  the SDSS images (see \secn{calculation}).
The mean value of the semi-major axis (sma) length for the outermost isophote for our sample 
galaxies is $5\arcsec.41$. The ellipse fitting process fails to 
converge before reaching the condition that isophotes have 
(mean isophotal intensity/total error in isophotal intensity) $\sim3$ for most of the cases,
principally because of crowding effects at the lowest brightness levels.
For the sample the mean signal-to-noise ratio reached at the outermost isophote is $14$. 
The average surface brightness and its average error at the outermost 
isophote we could reach for our sample
galaxies are $26.94\magarcsecsqi$ and  
$0.12\magarcsecsqi$ respectively. 

\begin{deluxetable}{lllllllllllll}
\tablecolumns{13}
\tablewidth{0pc}

\rotate
\setlength{\tabcolsep}{0.02in}
\tabletypesize{\scriptsize}
\tablecaption{Basic parametrs of the sample of early-type galaxies\labtablem{basic}}
\tablehead{
\colhead{}    &  \colhead{SDSS ID}&
\colhead{RA} & \colhead{Dec} & 
\colhead{{\it u}} & \colhead{{\it g}} & \colhead{{\it r}} & \colhead{{\it i}} & \colhead{{\it z}}&
\colhead{$R_{50}$}& \colhead{sma} & \colhead{$\mu_{{\it i}}$} & \colhead{SNR} \\

\colhead{}    &  \colhead{}&
\colhead{}  & \colhead{} & 
\colhead{$\unit{mag}$}& \colhead{$\unit{mag}$} & \colhead{$\unit{mag}$}& \colhead{$\unit{mag}$}& \colhead{$\unit{mag}$} &
\colhead{$\unit{arcsec}$}& \colhead{$\unit{arcsec}$} & \colhead{$\magarcsecsqi$}& \colhead{}\\

\colhead{(1)}    &  \colhead{(2)}&
\colhead{(3)}  & \colhead{(4)} & 
\colhead{(5)}& \colhead{(6)} & \colhead{(7)}& \colhead{(8)}& \colhead{(9)} &
\colhead{(10)}& \colhead{(11)} & \colhead{(12)}& \colhead{(13)}

}
\startdata
LFC 462   & 588848899913548358 & 182.02545140 &  -0.00014042 & 21.50 $\pm$ 0.43 & 21.06 $\pm$ 0.13 & 20.55 $\pm$ 0.13 & 20.57 $\pm$ 0.21 & 19.96 $\pm$ 0.43 & 1.07 $\pm$ 0.16 & 3.81 & 26.36 $\pm$ 0.06 & 17.25 \\ 
LFC 525   & 587748929241612656 & 182.10068456 &  0.00579127 & 22.48 $\pm$ 1.58 & 20.76 $\pm$ 0.09 & 20.42 $\pm$ 0.10 & 20.13 $\pm$ 0.12 & 20.74 $\pm$ 0.95 & 1.10 $\pm$ 0.11 & 3.81 & 26.06 $\pm$ 0.04 & 26.07 \\ 
LFC 554   & 587748929241547219 & 181.99182097 &  0.00889758 & 22.09 $\pm$ 1.04 & 21.89 $\pm$ 0.24 & 21.11 $\pm$ 0.18 & 19.90 $\pm$ 0.09 & 19.67 $\pm$ 0.34 & 1.28 $\pm$ 0.07 & 3.81 & 26.67 $\pm$ 0.08 & 12.50 \\ 
LFC 558   & 587748929241547253 & 181.99959155 &  0.01032991 & 26.27 $\pm$ 2.14 & 21.73 $\pm$ 0.24 & 20.20 $\pm$ 0.09 & 19.44 $\pm$ 0.07 & 19.05 $\pm$ 0.21 & 1.01 $\pm$ 0.07 & 5.11 & 26.58 $\pm$ 0.10 & 10.06 \\ 
LFC 622   & 587748929241546923 & 182.03581581 &  0.01761600 & 20.79 $\pm$ 0.41 & 19.26 $\pm$ 0.03 & 18.16 $\pm$ 0.02 & 17.70 $\pm$ 0.02 & 17.33 $\pm$ 0.06 & 1.38 $\pm$ 0.02 & 10.11 & 27.21 $\pm$ 0.09 & 11.88 \\ 
LFC 633   & 587748929241547046 & 182.04195384 &  0.01802068 & 20.48 $\pm$ 0.30 & 20.09 $\pm$ 0.06 & 19.51 $\pm$ 0.05 & 19.11 $\pm$ 0.05 & 19.21 $\pm$ 0.28 & 1.26 $\pm$ 0.05 & 8.32 & 27.18 $\pm$ 0.08 & 13.77 \\ 
LFC 663   & 587748929241612845 & 182.17111251 &  0.01778466 & 21.46 $\pm$ 0.60 & 21.97 $\pm$ 0.26 & 20.89 $\pm$ 0.15 & 20.06 $\pm$ 0.10 & 19.19 $\pm$ 0.22 & 1.14 $\pm$ 0.08 & 3.81 & 25.90 $\pm$ 0.05 & 23.48 \\ 
LFC 677   & 587748929241547621 & 181.99102394 &  0.02035316 & 19.94 $\pm$ 0.20 & 22.14 $\pm$ 0.42 & 20.94 $\pm$ 0.16 & 20.98 $\pm$ 0.33 & 22.79 $\pm$ 4.34 & 0.80 $\pm$ 0.17 & 5.36 & 27.64 $\pm$ 0.19 & 5.11 \\ 
LFC 725   & 587722983351583067 & 182.01578218 &  0.02274361 & 25.14 $\pm$ 4.14 & 21.58 $\pm$ 0.27 & 20.82 $\pm$ 0.13 & 20.75 $\pm$ 0.30 & 20.12 $\pm$ 0.77 & 1.32 $\pm$ 0.37 & 3.81 & 26.65 $\pm$ 0.12 & 8.92 \\ 
LFC 741   & 587748929241547216 & 181.99078515 &  0.02434438 & 21.46 $\pm$ 0.57 & 21.65 $\pm$ 0.19 & 20.74 $\pm$ 0.12 & 20.19 $\pm$ 0.12 & 19.80 $\pm$ 0.36 & 0.97 $\pm$ 0.09 & 4.86 & 27.60 $\pm$ 0.21 & 4.73 \\ 
LFC 824   & 587748929241547430 & 182.05600812 &  0.03170720 & 23.94 $\pm$ 10.03 & 23.14 $\pm$ 2.52 & 20.28 $\pm$ 0.18 & 20.44 $\pm$ 0.65 & 19.60 $\pm$ 0.78 & 0.74 $\pm$ 0.33 & 3.81 & 25.93 $\pm$ 0.04 & 28.48 \\ 
LFC 848   & 587748929241547282 & 182.00945210 &  0.03522617 & 22.05 $\pm$ 1.24 & 20.93 $\pm$ 0.14 & 20.26 $\pm$ 0.11 & 20.13 $\pm$ 0.17 & 19.63 $\pm$ 0.39 & 1.00 $\pm$ 0.12 & 3.81 & 26.89 $\pm$ 0.09 & 11.38 \\ 
LFC 925   & 587748929241612771 & 182.14531708 &  0.03984272 & 21.09 $\pm$ 0.52 & 21.00 $\pm$ 0.13 & 20.09 $\pm$ 0.07 & 19.76 $\pm$ 0.09 & 19.08 $\pm$ 0.24 & 1.27 $\pm$ 0.07 & 3.81 & 26.10 $\pm$ 0.05 & 21.72 \\ 
LFC 954   & 587748929241612750 & 182.13801353 &  0.04149656 & 24.88 $\pm$ 5.75 & 23.04 $\pm$ 0.87 & 20.93 $\pm$ 0.17 & 20.11 $\pm$ 0.14 & 20.28 $\pm$ 0.78 & 1.06 $\pm$ 0.15 & 3.63 & 26.12 $\pm$ 0.08 & 12.67 \\ 
LFC 985   & 587748929241547693 & 182.03813760 &  0.04718642 & 25.96 $\pm$ 2.24 & 25.08 $\pm$ 1.99 & 20.94 $\pm$ 0.14 & 19.70 $\pm$ 0.08 & 19.39 $\pm$ 0.25 & 1.21 $\pm$ 0.10 & 6.21 & 28.07 $\pm$ 0.20 & 4.93 \\ 
\enddata
\tablecomments{Here we list data for only the first 15 galaxies in our sample: the full table 
is available  at the end of the article as \tablem{basic full}. Column  (1) gives ID of the sample galaxies in SExtractor catalog generated
by us. Quantities from column (2) to column (10) are obtained from the 
 PhotoObjAll table of the SDSS DR7 photometric catalog \citep{aba09}.
Column (2) gives a unique SDSS identifier which is listed as objID in the PhotoObjAll table 
Columns (3) and (4) give J2000 right ascension and declination in degree. Columns (5) to (9)
give extinction corrected  Petrosian magnitudes in {\it u}, {\it g}, {\it r},{\it i} 
and {\it z} bands. Column (10) gives Petrosian half-light radius in a arcsec.
Columns (11) to (13) give length of semi-major axis (in arcsec), {\it i}-band surface brightness
(in $mag/arcsec^{-2}$) and signal to noise ratio respectively of the outermost isophote in which
ellipse can be fitted (see Section 4).
}
\end{deluxetable}

\subsection{Bulge-disk decomposition}
\labsubsecn{decomposition}

We have performed
bulge-disk decomposition and selected 132 galaxies with
bulge-to-total luminosity ratio $B/T >0.4$ as being of early-type \citep[see][for a discussion]{raw07}, 
which will form 
the basis for work reported in this paper. Here the luminosities are
estimated using parameters which define the large scale
structure of the galaxy. We assume that  each galaxy  consists
of a bulge and a disk, and used the two-dimensional code  \texttt{GALFIT} \citep{pen02} 
to decompose the image of the galaxy into these components.
The intensity profile of the bulge is modeled with the Sersic law \citep{ser68},
\begin {equation}
\Sigma(r)=\Sigma_e \mbox{exp}\bbbs{-2.303b_{n} \bbbs\bbb{\frac{r}{r_e}}\ebb^{1/n} - 1\ebbs}\ebbs,
\labequn{sersic}
\end {equation}
where $r_{e}$ is the half light radius, $\Sigma_e$ the surface brightness at $r_{e}$ and $n$ 
the Sersic index. The disk is modeled with an exponential profile \citep{fre70}) 
\begin {equation}
\Sigma(r)=\Sigma_0 \mbox{exp}\bbbs{-\frac{r}{r_d}}\ebbs,
\labequn{exponential}
\end {equation}
where $\Sigma_0$ is the central surface brightness and $r_{d}$ is the disk scale length.  
\texttt{GALFIT} obtains best fit parameters by minimizing
$\chi^2$ obtained by comparing the observed distribution of intensity in the 
two-dimensional image of the galaxy with model images obtained using the radial profiles 
defined above, and convolved with the point spread
function. After the decomposition, the total bulge and disk fluxes 
are obtained by integrating the best fit profiles over all values of the 
semi-major axis. 

\subsection{Spectroscopic data}
\labsubsecn{spectroscopic}

The distances to most of our galaxies are unknown. However, we
have obtained spectra for some of the galaxies in the LFC 1208+0010 field
using the 2dF/AAOmega multi-fiber system on the AAT. For these galaxies we
can derive distances, absolute magnitudes and linear sizes, to facilitate
comparison with other samples. These single-fiber spectra do not sample
our galaxies consistently and were only intended for determining
redshifts, but they do confirm that most of our early-type galaxies
selected on the basis of B/T are indeed early-type galaxies.

We chose to observe the 266 selected galaxies from the field 
SDSS 1208+0010 (see \subsecn{sample}) using the AAOmega multi-fiber 
spectrograph\footnote{See: http://www.aao.gov.au/local/www/aaomega/} 
\citep{sau04} on the $4\mtr$ Anglo-Australian 
Telescope (AAT).  AAOmega, which has 392 fibers each with $2\arcsec$ 
angular diameter, was used with gratings
385R and 580V in multi-object mode. The detectors used were 
EEV 4482 $2\rm{k}\times4\rm{k}$ CCDs having pixel size $15\,\micron$,
allowing 10 pixels per fiber. The wavelength range covered by 
grating 580V is $3700 - 5700\angstrom$ with dispersion $1.0\angstrompixi$,
and by grating 385R is $5600 - 8800\angstrom$ with 
dispersion $1.6\angstrompixi$. This setup of AAOmega
provides a resolution of $R\sim1300$.  

The compactness of our field as compared to the $2\unit{degree}$ 
field of view of AAOmega, together with a $30\arcsec$ restriction on 
the minimum separation between fibers, meant that
it was not possible to observe all the 266 galaxies simultaneously. 
We used the latest version of the AAOmega specific {\it CONFIGURE} 
program\footnote{See: http://www.aao.gov.au/AAO/2df/aaomega/aaomega\_software.html}
for automatic fiber allocation. This program uses a ``simulated annealing'' 
algorithm \citep{mis06} to maximize the number of objects observed. We assigned 
priorities to our targets as a function of fiber magnitude in 
the {\it g} filter ({\it fiberMag\_g}). We selected the positions of blank sky 
regions and suitable guide stars from the SDSS {\it PhotoObjAll} 
table, lying within the 2dF field but outside the central LFC field. 
SDSS images were visually inspected to ensure 
there was no contamination by other objects. 

The publicly available 
{\it 2DFDR}\footnote{http://www.aao.gov.au/AAO/2df/software.html\#2dfdr} 
\citep{bai04} pipeline reduction system was used
on each exposure to subtract bias and dark current and 
to carry out flat-fielding, sky-subtraction and wavelength-calibration.
The blue and red-arm spectra were spliced together after the spectra in each exposure set
were co-added.
Redshifts were determined using a version of the {\it ZCODE} software in manual 
mode \citep[see][sec 5.2 \& 5.3]{can06}. Each spectrum was visually
inspected and assigned a redshift quality flag $Q$, having value
from 0 to 6 based on (i) emission-line/absorption-line features identified in the 
spectrum and (ii) cross-correlation value estimated by the software. 
The cross-correlation was done with that template spectrum which matched 
best with each observed spectrum. A larger value of $Q$ signifies a more reliable 
redshift measurement. Redshift measurements with quality $Q\geq3$  were considered
reliable,  while $Q <3$ was assigned to unreliable redshifts or when
no redshift was obtained. From experience with much larger samples of
galaxies for other projects  \citep[e.g.][]{can06}, we believe that around 95\% of 
redshifts flagged as reliable should be correct.  

The median Petrosian half-light radius of our sample of galaxies is about
$1.05\arcsec$, only slightly larger than the radius of the 2dF/AAOmega
fibers. Thus for large nearby galaxies our spectra sample only the
nucleus, while for small distant galaxies we integrate most of the
light. The majority of our spectra were best fitted by an early-type galaxy
template, confirming the validity of our selection criteria based on
$B/T$. We defer further discussion of the spectra to a subsequent
paper: here we simply use the spectra to determine redshifts and hence
distances.

AAOmega observations for galaxies in field SDSS 1208 were carried out in 
service mode with exposures of $1\unit{h}$ ($3\times1200\unit{sec}$) in June 2006 and 
for $1.5\unit{h}$ ($3\times1800\unit{sec}$) in March 2007. 
In the observations of June 2006 only 135 galaxies could be 
targeted. We attained a mean S/N per pixel of $\sim4$ in the 
blue and $\sim10$ in the red spectra for targets with $\rm{{\it fiberMag\_g}}\sim21$. 
We obtained reliable redshifts for 55/135
(40\%) of the galaxies targeted and achieved 100\% success for the brightest
targets, i.e. for all 37 galaxies with $\rm{{\it fiberMag\_g}} < 21.2$.  
However, the success rate fell rapidly for fainter galaxies.  
We observed the field SDSS 1208 again in March 2007 with increased exposure time 
and could target 121 galaxies from the field, including all those not targetted 
in 2006 plus 58 that failed to yield a redshift in the first run.
By combining the spectra from the two observing runs 
we determined reliable redshifts for an additional 17 galaxies.
It should be noted that our success in determining redshifts is higher when
emission lines are present, and that our fiber spectra sample only a $2\arcsec$
diameter patch of sky which may contain a substantial fraction of
the light from small distant galaxies but only the core of nearby
galaxies. Thus they do not give an unbiased sampling of the full set of
galaxies or of their spectral types.

We show in \fig{redshift} the redshift distribution for all the
galaxies in field SDSS 1208 for which we obtained redshifts. The
redshifts of galaxies which we selected as early-type on the basis of $B/T$ are
shown by the shaded area. It is evident that the proportion of
early-type galaxies increases with redshift, with late-type galaxies
being more common for $z<0.2$ while early-type galaxies are more common for
$z>0.2$. 

We have obtained the rest frame absolute magnitudes in 
the {\it u}, {\it g}, {\it r}, {\it i} and {\it z}
bands for the galaxies for which we have redshift measurements using the equation

\begin{equation}
M = m - DM(z) - k(z) - A,
\labequn{absolute magnitude}
\end{equation}
where $z$ is the redshift, $\appm{}$ and $\absm{}$  are 
apparent and absolute magnitudes respectively in each band and $k(z)$ and $A$ 
are the k-correction and Galactic extinction in that band. The distance 
modulus $DM(z)$ is obtained using  a flat $\lambda$CDM
cosmology with $\Omega_{m}=0.3$, $\Omega_{\Lambda}=0.7$ and
 $H_{0}=70\hubbleunit$. The k-corrections have been 
obtained  using the publicly available KCORRECT code
\citep{bla07}, and the extinction values are obtained from  the 
DR7 photometric catalog which uses Galactic extinction maps published by 
\citet{sch98}.
 
To compare our results with earlier work,  we have obtained Johnson 
{\it B} and {\it V} magnitudes using the transformations \citep{smi02} 
\begin{eqnarray}
 B = g + 0.47 (g - r) + 0.17 \nonumber \\
 V = g - 0.55 (g -r) - 0.03.
\labequn{trans}
\end{eqnarray}
The apparent {\it B, V} magnitudes are converted to absolute magnitudes using 
\equn{absolute magnitude} and
k-corrections in {\it B} and {\it V} bands are obtained from \citet{pog97}. 

To summarize, from the two sets of AAT observations we obtained reliable redshifts 
for 129 of the 266 galaxies in our original photometric sample: of these, 53 galaxies are 
included in the early-type sample discussed in this paper. The redshift-dependent
parameters are listed in the \tablem{redshift_dep}.
 In order to facilitate 
comparison with other samples of galaxies, plots of absolute {\it B} magnitude 
and Petrosian $R_{50}$ radius against redshift are shown in \fig{absmag}. 
The bulk of the galaxies have redshifts in the range $0.1 < z < 0.3$, 
with $\absm{B}$ between -17 and -21 and Petrosian $R_{50}$ from two to six kpc. There are half a 
dozen dwarfs with $z < 0.1$ and a similar number of giants, mostly at high redshifts. The upper 
panel of \fig{absmag} shows a good correlation between Petrosian $R_{50}$ and $z$ for most of 
the galaxies.

\begin{figure}
        \centering
\includegraphics[width=1.0\textwidth,angle=0.0,clip]{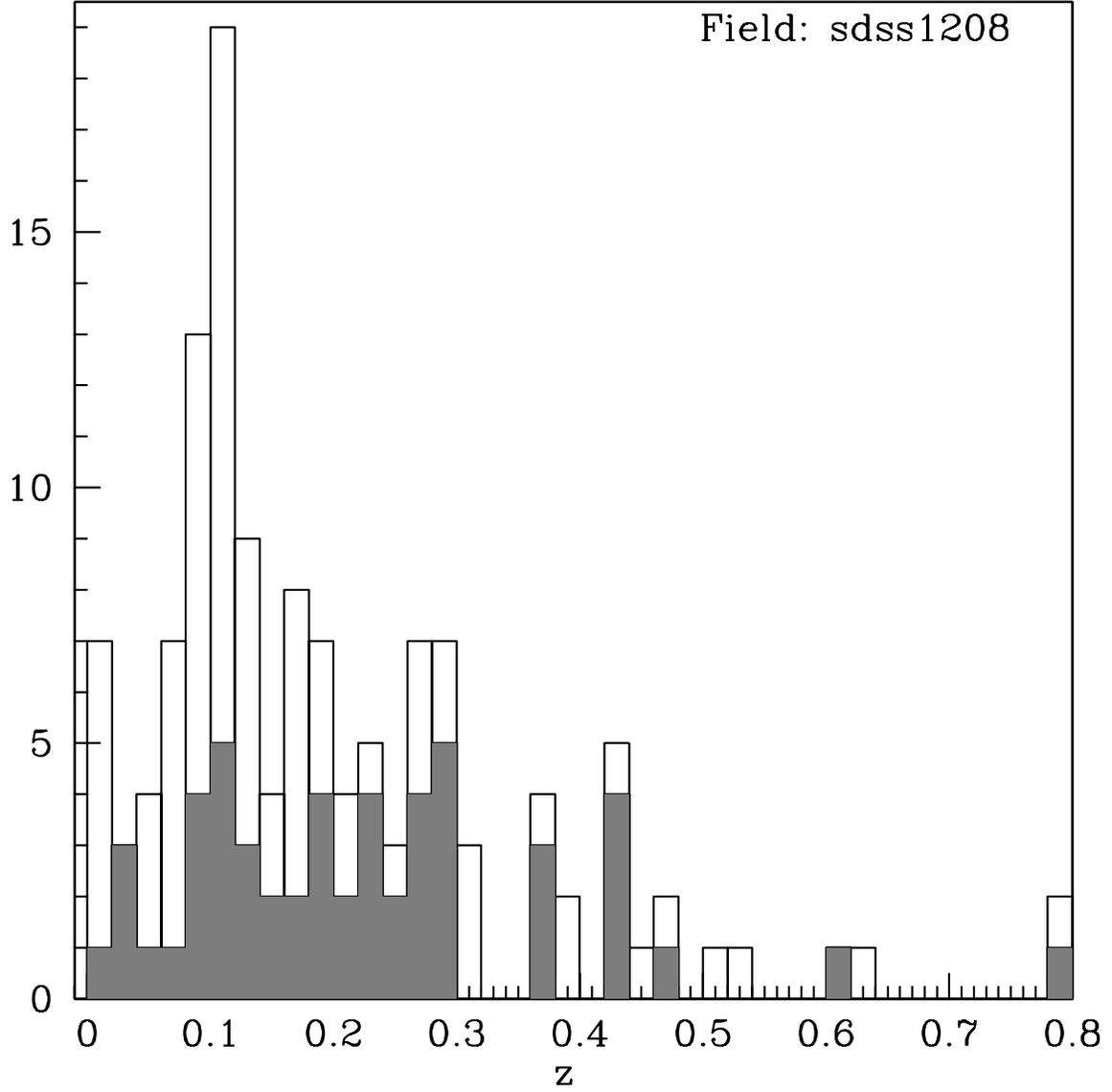}

\caption{Redshift distribution of the 129 galaxies from field sdss1208. The shaded 
area shows the distribution of 53 early type galaxies which are part of this work.}
\labfig{redshift}
\end{figure}

\begin{figure}
        \centering
\includegraphics[width=1.0\textwidth,angle=0.0,clip]{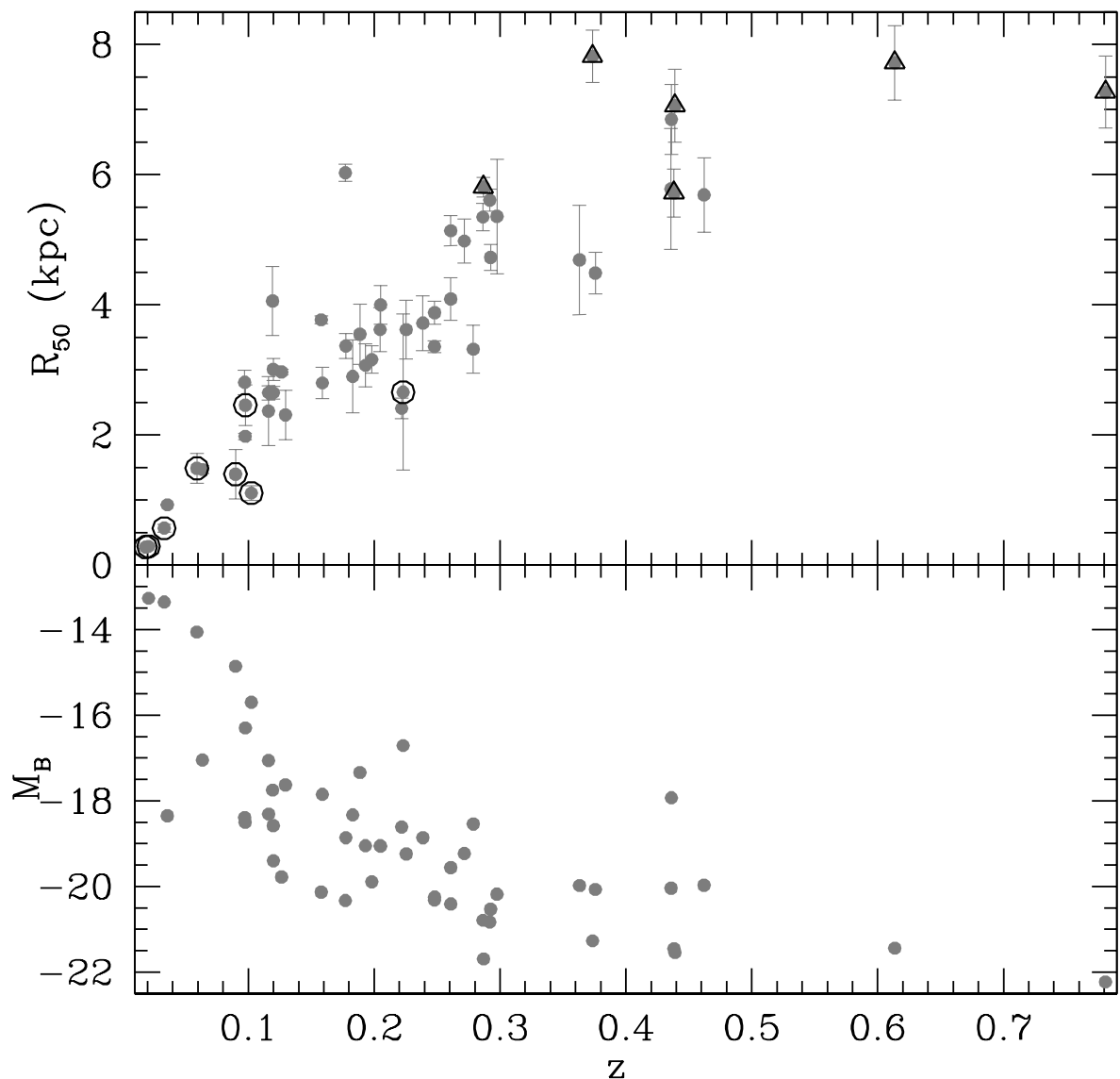}

\caption{ The lower panel show plot of redshift versus $\absm{B}$
for the 53 galaxies of our sample. See text for details. 
The lowest redshift galaxies are mostly dwarfs, while those in the high 
redshift tail are the brightest. The plot of
redshift versus Petrosian $R_{50}$ radius in the upper panel shows a
good correlation. The dwarf galaxies (with $\absm{B} > -17$) are indicated by
large circles, and the most luminous ($\absm{B} < -21$) with traingle symbols.}
\labfig{absmag}
\end{figure}

\begin{deluxetable}{lllllllllll}
\tablecolumns{11}
\tablewidth{0pc}

\tabletypesize{\footnotesize}

\tablecaption{Distance-dependent parameters for 53 galaxies with
  redshifts
\labtablem{redshift_dep}}
\tablehead{
\colhead{ID}    &  \colhead{z}&
\colhead{$R_{50}$} & \colhead{$sma$} & 
\colhead{$\absm{\it u}$} & \colhead{$\absm{\it g}$} & \colhead{$\absm{\it r}$} & 
\colhead{$\absm{\it i}$} & \colhead{$\absm{\it z}$} &   
\colhead{$\absm{B}$} & \colhead{$\absm{V}$}   \\

\colhead{}    &  \colhead{}&
\colhead{$\kpc$} & \colhead{$\kpc$} & 
\colhead{$\unit{mag}$}& \colhead{$\unit{mag}$} & \colhead{$\unit{mag}$} & \colhead{$\unit{mag}$} & \colhead{$\unit{mag}$} &
 \colhead{$\unit{mag}$} & \colhead{$\unit{mag}$}  \\

\colhead{(1)}    &  \colhead{(2)}&
\colhead{(3)} & \colhead{(4)} & 
\colhead{(5)}& \colhead{(6)} & \colhead{(7)} & \colhead{(8)} & \colhead{(9)} &
 \colhead{(10)} & \colhead{(11)} 
}
\startdata
LFC 558   & 0.4382 & 5.72 $\pm$ 0.37 & 28.97 &	-18.48 & -21.90 & -22.49 & -22.85 & -23.07 & 	-21.45 & -22.26 \\ 
LFC 622   & 0.1580 & 3.77 $\pm$ 0.06 & 27.59 &	-19.24 & -20.66 & -21.43 & -21.78 & -22.11 & 	-20.13 & -21.12 \\ 
LFC 633   & 0.1162 & 2.65 $\pm$ 0.11 & 17.50 &	-18.26 & -18.73 & -19.27 & -19.54 & -19.47 & 	-18.31 & -19.06 \\ 
LFC 663   & 0.6136 & 7.72 $\pm$ 0.57 & 25.75 &	-21.39 & -22.18 & -23.40 & -23.79 & -24.52 & 	-21.44 & -22.88 \\ 
LFC 824   & 0.2230 & 2.66 $\pm$ 1.20 & 13.68 &	-17.42 & -17.94 & -20.20 & -19.88 & -20.67 & 	-16.71 & -19.21 \\ 
LFC 848   & 0.2255 & 3.62 $\pm$ 0.45 & 13.80 &	-18.46 & -19.60 & -20.01 & -20.02 & -20.48 & 	-19.24 & -19.86 \\ 
LFC 985   & 0.4361 & 6.85 $\pm$ 0.54 & 35.11 &	-19.53 & -19.47 & -22.37 & -23.17 & -23.21 & 	-17.93 & -21.10 \\ 
LFC 988   & 0.2788 & 3.32 $\pm$ 0.37 & 16.13 &	-17.20 & -19.09 & -19.92 & -20.69 & -21.08 & 	-18.54 & -19.58 \\ 
LFC 1039  & 0.3632 & 4.69 $\pm$ 0.84 & 29.91 &	-19.70 & -20.37 & -20.85 & -20.52 & -18.35 & 	-19.98 & -20.66 \\ 
LFC 1056  & 0.0636 & 1.47 $\pm$ 0.06 & 9.70 &	-16.05 & -17.47 & -18.02 & -18.18 & -18.19 & 	-17.05 & -17.81 \\ 
LFC 1091  & 0.4359 & 5.78 $\pm$ 0.93 & 28.88 &	-18.56 & -20.78 & -22.00 & -22.42 & -22.94 & 	-20.04 & -21.48 \\ 
LFC 1411  & 0.1588 & 2.80 $\pm$ 0.24 & 10.45 &	-17.72 & -18.35 & -19.06 & -19.16 & -19.15 & 	-17.85 & -18.77 \\ 
LFC 1494  & 0.2919 & 5.61 $\pm$ 0.17 & 36.36 &	-20.60 & -21.36 & -22.13 & -22.61 & -22.69 & 	-20.83 & -21.81 \\ 
LFC 1504  & 0.2716 & 4.98 $\pm$ 0.34 & 27.07 &	-18.55 & -19.75 & -20.50 & -21.05 & -20.73 & 	-19.23 & -20.19 \\ 
LFC 1551  & 0.1829 & 2.90 $\pm$ 0.56 & 11.72 &	-19.07 & -18.93 & -19.85 & -20.02 & -21.28 & 	-18.33 & -19.47 \\ 
LFC 1613  & 0.1161 & 2.37 $\pm$ 0.53 & 11.27 &	-17.10 & -17.47 & -17.99 & -17.76 & -17.65 & 	-17.06 & -17.79 \\ 
LFC 1739  & 0.0332 & 0.57 $\pm$ 0.06 & 3.38 &	-9.69 & -13.95 & -14.83 & -15.53 & -16.20 & 	-13.36 & -14.46 \\ 
LFC 1772  & 0.1775 & 3.37 $\pm$ 0.19 & 16.90 &	-16.78 & -19.34 & -20.00 & -20.00 & -20.20 & 	-18.86 & -19.73 \\ 
LFC 1802  & 0.2925 & 4.73 $\pm$ 0.20 & 25.88 &	-19.77 & -20.89 & -21.30 & -21.50 & -21.32 & 	-20.53 & -21.15 \\ 
LFC 2159  & 0.2387 & 3.72 $\pm$ 0.42 & 14.40 &	-18.33 & -19.53 & -20.60 & -20.95 & -22.08 & 	-18.86 & -20.15 \\ 
LFC 2168  & 0.2480 & 3.88 $\pm$ 0.18 & 14.82 &	-19.68 & -20.59 & -20.94 & -20.76 & -20.94 & 	-20.25 & -20.81 \\ 
LFC 2177  & 0.2047 & 3.62 $\pm$ 0.34 & 17.16 &	-18.26 & -19.50 & -20.12 & -19.98 & -20.39 & 	-19.05 & -19.87 \\ 
LFC 2180  & 0.1024 & 1.11 $\pm$ 0.11 & 6.77 &	-16.46 & -16.19 & -16.87 & -17.71 & -18.57 & 	-15.70 & -16.59 \\ 
LFC 2181  & 0.2479 & 3.36 $\pm$ 0.09 & 19.84 &	-19.46 & -20.74 & -21.28 & -21.46 & -21.78 & 	-20.32 & -21.07 \\ 
LFC 2184  & 0.0357 & 0.93 $\pm$ 0.01 & 7.92 &	-18.01 & -18.64 & -18.91 & -18.97 & -19.04 & 	-18.35 & -18.82 \\ 
LFC 2310  & 0.1931 & 3.07 $\pm$ 0.33 & 12.24 &	-19.22 & -19.47 & -20.02 & -20.10 & -20.61 & 	-19.05 & -19.80 \\ 
LFC 2356  & 0.1193 & 4.06 $\pm$ 0.53 & 8.20 &	-16.73 & -18.07 & -18.40 & -18.45 & -18.74 & 	-17.75 & -18.28 \\ 
LFC 2362  & 0.0592 & 1.49 $\pm$ 0.23 & 7.84 &	-15.51 & -14.99 & -16.61 & -17.28 & -17.74 & 	-14.06 & -15.91 \\ 
LFC 2405  & 0.0899 & 1.40 $\pm$ 0.38 & 5.80 &	-12.28 & -15.57 & -16.73 & -16.86 & -17.96 & 	-14.86 & -16.24 \\ 
LFC 2414  & 0.2869 & 5.81 $\pm$ 0.15 & 37.72 &	-21.21 & -22.22 & -22.97 & -23.28 & -23.56 & 	-21.69 & -22.66 \\ 
LFC 2484  & 0.1887 & 3.55 $\pm$ 0.46 & 13.91 &	-19.48 & -18.08 & -19.31 & -19.39 & -19.11 & 	-17.34 & -18.79 \\ 
LFC 2525  & 0.2051 & 4.00 $\pm$ 0.30 & 12.83 &	-18.26 & -19.44 & -19.90 & -20.07 & -20.00 & 	-19.06 & -19.72 \\ 
LFC 2549  & 0.1264 & 2.97 $\pm$ 0.04 & 21.78 &	-19.02 & -20.26 & -20.93 & -21.24 & -21.49 & 	-19.78 & -20.66 \\ 
LFC 2692  & 0.1295 & 2.31 $\pm$ 0.38 & 11.79 &	-14.39 & -17.83 & -17.90 & -17.70 & -19.00 & 	-17.63 & -17.90 \\ 
LFC 2708  & 0.0184 & 0.28 $\pm$ 0.02 & 1.42 &	-8.13 & -12.93 & -14.19 & -14.57 & -12.69 & 	-12.17 & -13.65 \\ 
LFC 2712  & 0.1198 & 3.01 $\pm$ 0.17 & 16.30 &	-17.70 & -19.21 & -20.21 & -20.74 & -21.25 & 	-18.58 & -19.79 \\ 
LFC 2782  & 0.1772 & 6.03 $\pm$ 0.13 & 40.61 &	-19.58 & -20.83 & -21.52 & -21.87 & -22.26 & 	-20.33 & -21.24 \\ 
LFC 3072  & 0.4621 & 5.69 $\pm$ 0.57 & 29.86 &	-20.85 & -20.65 & -21.73 & -21.94 & -22.48 & 	-19.97 & -21.28 \\ 
LFC 3094  & 0.2864 & 5.35 $\pm$ 0.21 & 32.55 &	-20.33 & -21.19 & -21.68 & -21.80 & -22.06 & 	-20.79 & -21.49 \\ 
LFC 3218  & 0.2218 & 2.41 $\pm$ 0.16 & 16.56 &	-18.37 & -19.02 & -19.53 & -19.74 & -19.92 & 	-18.61 & -19.33 \\ 
LFC 3274  & 0.0977 & 2.46 $\pm$ 0.31 & 6.56 &	-16.43 & -16.74 & -17.31 & -17.69 & -17.57 & 	-16.30 & -17.08 \\ 
LFC 3292  & 0.4390 & 7.06 $\pm$ 0.56 & 35.24 &	-20.69 & -21.76 & -21.85 & -22.06 & -21.94 & 	-21.54 & -21.84 \\ 
LFC 3381  & 0.7809 & 7.27 $\pm$ 0.55 & 38.03 &	-22.38 & -22.79 & -23.63 & -23.38 & -23.31 & 	-22.23 & -23.28 \\ 
LFC 3430  & 0.1200 & 2.65 $\pm$ 0.10 & 24.12 &	-18.87 & -19.89 & -20.58 & -20.91 & -21.26 & 	-19.40 & -20.30 \\ 
LFC 3458  & 0.3757 & 4.49 $\pm$ 0.32 & 25.14 &	-20.25 & -20.69 & -21.65 & -22.01 & -22.50 & 	-20.07 & -21.25 \\ 
LFC 3508  & 0.2609 & 5.14 $\pm$ 0.23 & 35.25 &	-19.75 & -20.97 & -21.82 & -22.27 & -22.49 & 	-20.41 & -21.47 \\ 
LFC 4009  & 0.1981 & 3.16 $\pm$ 0.21 & 20.33 &	-19.19 & -20.31 & -20.84 & -21.12 & -21.49 & 	-19.89 & -20.63 \\ 
LFC 4059  & 0.3735 & 7.82 $\pm$ 0.40 & 52.06 &	-19.50 & -21.67 & -22.17 & -22.43 & -22.36 & 	-21.27 & -21.98 \\ 
LFC 4188  & 0.0975 & 1.98 $\pm$ 0.05 & 10.66 &	-17.79 & -18.99 & -19.69 & -19.97 & -20.15 & 	-18.50 & -19.41 \\ 
LFC 4248  & 0.0972 & 2.81 $\pm$ 0.19 & 14.96 &	-17.78 & -18.73 & -19.10 & -19.24 & -19.55 & 	-18.39 & -18.96 \\ 
LFC 4394  & 0.2608 & 4.09 $\pm$ 0.33 & 15.37 &	-17.94 & -19.84 & -20.09 & -20.33 & -20.61 & 	-19.56 & -20.01 \\ 
LFC 4536  & 0.2975 & 5.36 $\pm$ 0.88 & 16.88 &	-19.52 & -20.46 & -20.68 & -21.02 & -21.17 & 	-20.18 & -20.61 \\ 
LFC 4763  & 0.0208 & 0.29 $\pm$ 0.01 & 1.84 &	-13.36 & -13.86 & -14.75 & -15.22 & -15.62 & 	-13.27 & -14.38 \\ 
\enddata
\tablecomments{ Column (1) gives our SExtractor catalog ID. Column (2) gives spectroscopic redshift
obtained by us. Column (3) and (4) give Petrosian half light radius and semi-major axis lenght 
in kiloparsec. Columns (5) to (11) give absolute magnitudes of galaxies in SDSS filters. Column (10) and 
and (11) give Johnson {\it B} and {\it V}  absolute magnitudes obtained using \dequn{absolute magnitude}{trans}.
}
\end{deluxetable}

\section{Isophotal analysis and estimation of parameters}
\labsecn{parameter estimation}

The early-type galaxies selected for our study are bulge dominated and 
therefore have isophotes which are close to elliptical in shape. It
is our aim to measure morphological parameters which describe the elliptical
shape, as well as small deviations from it, as a function of galaxy properties.
We performed ellipse fitting 
using the IRAF routine \texttt{ellipse} within STSDAS, which is based on a
technique described by  \citet{jed87}.  However, before the ellipse 
fitting, pre-processing using SExtractor has to be performed to
obtain a catalog of all detected objects in the field, 
a sky background image, and a {\it segmentation } image.
 
The segmentation image is a map of all detected objects in the field
in which, for any given object, all the pixels have value 
equal to the running SExtractor catalog number corresponding to that object.
The value of parameters used for estimation of
background in the SExtractor configuration file are BACK\_SIZE 128; 
BACK\_FILTERSIZE 3; BACKPHOTO\_TYPE LOCAL; and BACKPHOTO\_THICK 64 
\citep[see][]{ber96}. The sky background
image obtained is subtracted from the field image, and 
$150\times150$ pixel cutouts around the geometric center 
of our sample galaxies are made from
the sky-subtracted image and the segmentation image.
The cut-outs of segmentation images are 
used to generate mask files for the surface photometry.

We use the geometric center, ellipticity  and position angle of 
sample galaxies, obtained from the SExtractor catalog, as initial values 
in the ellipse fitting. 
In the \texttt{ellipse} task, the image intensity is first sampled along a 
trial ellipse generated using these parameters, and the intensity string
$I(\theta)$ is expanded in a Fourier series, 
\begin{equation}
I(\theta) = I_{0} + \sum_{n=1}^N \bbs A_{n} \sin(n\theta) + 
 B_{n} \cos(n\theta)\ebs, 
\labequn{ellipse}
\end{equation} 
where $N$ is the highest harmonic fitted and $\theta $ is the azimuthal 
angle measured from the major axis.  \texttt{ellipse} uses the  first and 
second order coefficients obtained for trial ellipses to
iteratively improve the fitting.  For perfectly elliptical isophotes 
all the Fourier coefficients except $I_{0}$, which is the mean isophotal 
intensity, should vanish at the end of 
the fitting process, and 
significant residuals can only be of order three or higher. 
We allowed the geometric center, ellipticity and position 
angle to vary freely during fitting. Successive ellipses are fitted along the 
semi-major axis with a logarithmic step of 0.05 until the ratio
(mean isophotal intensity/total 
error in isophotal intensity) $\sim 3$, or the ellipse fitting
process fails to converge. The output of  \texttt{ellipse} is a table 
containing radial profiles of number of
isophotal parameters along with uncertainties associated with them 
\citep[see][for error estimation of isophotal parameters]{Busko1996}.

In their seminal papers on the isophotal 
shapes of elliptical galaxies, \citet{ben88} and \citet{ben89} define 
structural parameter $a_{n}/a$ and $b_{n}/a$.
The $A_{n}$ and $B_{n}$ provided  by the  \texttt{ellipse} table 
are normalized to the semi-major axis $a$ and the local intensity gradient 
intensity $\deri{I}{a}$ (\texttt{ellipse} output GRAD). 
The normalized $B_{n}$ coefficients can be converted to the $a_{n}/a$ parameter 
as 
\begin{equation}
\frac{a_{n}}{a} = B_{n}\sqrt{1 - \epsilon} = B_{n}\sqrt{b/a},
\labequn{B4 to a4}
\end{equation}
where $b$ is the semi-minor axis 
and  $\epsilon$ is isophotal ellipticity \citep{mil99,tre07,hao06b}.
The factor of $\sqrt{1 - \epsilon} = \sqrt{b/a}$ is needed to 
re-normalize the radial deviation to the equivalent radius
$r = \sqrt{ab}$ \citep{mil99}. The $a_{4}/a$ parameter, which 
is the most dominant coefficient for any isophote deviating 
from a pure ellipse, quantifies the deviation 
along the major axis of the isophote. Isophotes 
with $a_{4}/a < 0$ have a ``boxy'' shape while the shape of isophotes 
with $a_{4}/a > 0$ is ``disky'' \citep{ben88}. The parameter $b_{n}/a$
can be related to the normalized $A_{n}$  parameter through a relation
similar to \equn{B4 to a4} but we do not use it in our present work.

The Fourier coefficients and morphological parameters  vary along the 
semi-major axis, and we find that there are significant changes in values
in the outer regions of the galaxies.  It is therefore not
enough to consider characteristic values of the parameters at
a single fiducial distance from the centre, as has been done by
earlier workers (see \eg \citet{ben88}, \citet{hao06}), for investigating 
relationships between these parameters and other global galaxy properties.
However, in comparing data for a large sample of galaxies it is convenient 
to derive mean values of isophote parameters in a few regions, rather than 
the full sets of isophotes.
 We therefore divide each galaxy into four regions defined
by distance from the centre along the semi-major axis: 
(i) seeing radius $r_{s}$ to $1.5R_{50}$ (Region 1), 
(ii) $1.5R_{50}$ to $3.0R_{50}$ (Region 2), 
(iii) $3.0R_{50}$ to $4.5R_{50}$ (Region 3) and 
(iv) semi-major axis $> 4.5R_{50}$ (Region 4), 
where $R_{50}$ is the Petrosian half-light radius. 
In each region we obtain the mean value of various parameters, 
weighted with the intensity (counts) and inversely weighted  
with the variance of the parameter. 
For example, the mean value of $a_{4}/a$ in Region 1 is obtained as  
\begin{equation}
\mean{\frac{a_{4}}{a}} = 
\frac{\displaystyle\int_{r_{s}}^{1.5R_{50}} 
\frac{a_{4}}{a}(r) I(r) \bbs \sigma_{\frac{a_{4}}{a}}(r)  \ebs^{-2} dr
}{\displaystyle\int_{r_{s}}^{1.5R_{50}} I(r) \bbs \sigma_{\frac{a_{4}}{a}}(r)  \ebs^{-2} dr    }
\labequn{average a4}
\end{equation}
Note that not all galaxies have
data in all four regions: for example, Region 1 does not exist if 
seeing radius $r_{s} > 1.5R_{50}$, while sometimes it is impossible to fit reliable
ellipses in regions 3 and 4 due to crowding and/or low $S/N$.
Some of the derived parameters for our sample of galaxies are listed
in \tablem{derived} (only the first 5 galaxies are listed here as
examples; the complete table is at the end of the article as \tablem{derived full}).

\begin{deluxetable}{llrrr}
\tablecolumns{5}
\tablewidth{0pc}

\tabletypesize{\footnotesize}

\tablecaption{Derived parameters for a sample of the LFC galaxies\labtablem{derived}}
\tablehead{
\colhead{ID}    &  \colhead{B/T}&
\colhead{$\mean{\frac{a3}{a}}$} & \colhead{$\mean{\frac{a4}{a}}$} & 
\colhead{$\mean{\epsilon}$} \\

\colhead{(1)}    &  \colhead{(2)}&
\colhead{(3)} & \colhead{(4)} & 
\colhead{(5)}

}
\startdata
LFC 462   & 0.89 & 	\nodata & \nodata & \nodata \\ 
&& 3.52e-03 $\pm$ 2.25e-03 & -1.25e-03  $\pm$  2.16e-03 & 0.057  $\pm$  0.0058 \\ 
&& 7.01e-03 $\pm$ 4.73e-03 & 6.91e-03  $\pm$  3.09e-03 & 0.048  $\pm$  0.0063 \\ 
&& \nodata & \nodata & \nodata \\ 
 
LFC 525   & 0.97 & 	\nodata & \nodata & \nodata \\ 
&& 5.52e-03 $\pm$ 1.89e-03 & -1.09e-03  $\pm$  1.53e-03 & 0.070  $\pm$  0.0047 \\ 
&& -6.61e-03 $\pm$ 5.70e-03 & 1.21e-02  $\pm$  5.41e-03 & 0.023  $\pm$  0.0117 \\ 
&& \nodata & \nodata & \nodata \\ 
 
LFC 554   & 0.65 & 	-4.29e-03 $\pm$ 1.96e-03 & 1.00e-02  $\pm$  1.82e-03 & 0.094  $\pm$  0.0040 \\ 
&& -8.35e-03 $\pm$ 3.77e-03 & 3.49e-03  $\pm$  3.75e-03 & 0.115  $\pm$  0.0071 \\ 
&& \nodata & \nodata & \nodata \\ 
&& \nodata & \nodata & \nodata \\ 
 
LFC 558   & 0.68 & 	6.77e-03 $\pm$ 1.47e-03 & 1.88e-03  $\pm$  1.42e-03 & 0.038  $\pm$  0.0035 \\ 
&& 4.92e-03 $\pm$ 1.60e-03 & 1.12e-04  $\pm$  1.56e-03 & 0.081  $\pm$  0.0036 \\ 
&& -1.10e-02 $\pm$ 1.19e-02 & -4.33e-03  $\pm$  1.28e-02 & 0.164  $\pm$  0.0248 \\ 
&& \nodata & \nodata & \nodata \\ 
 
LFC 622   & 0.59 & 	-4.88e-04 $\pm$ 1.06e-03 & 1.51e-02  $\pm$  4.42e-04 & 0.358  $\pm$  0.0018 \\ 
&& -1.18e-03 $\pm$ 8.45e-04 & 1.29e-02  $\pm$  4.69e-04 & 0.421  $\pm$  0.0013 \\ 
&& 4.78e-02 $\pm$ 4.35e-03 & -1.47e-02  $\pm$  3.35e-03 & 0.310  $\pm$  0.0075 \\ 
&& 3.79e-02 $\pm$ 1.57e-02 & -2.56e-02  $\pm$  1.45e-02 & 0.259  $\pm$  0.0257 \\ 
 
LFC 633   & 0.66 & 	5.30e-03 $\pm$ 7.48e-04 & -2.57e-03  $\pm$  6.74e-04 & 0.213  $\pm$  0.0016 \\ 
&& 1.59e-02 $\pm$ 1.24e-03 & -3.83e-03  $\pm$  1.11e-03 & 0.219  $\pm$  0.0027 \\ 
&& 1.64e-02 $\pm$ 4.29e-03 & -6.56e-04  $\pm$  4.22e-03 & 0.269  $\pm$  0.0074 \\ 
&& -2.51e-01 $\pm$ 7.91e-02 & -2.39e-02  $\pm$  2.74e-02 & 0.102  $\pm$  0.0591 \\

\enddata
\tablecomments{ Here we list data for only the first five galaxies in
  our sample: the full table is at the end of the article as \tablem{derived full}.  Column (1)
  gives our SExtractor catalog ID. Column (2) gives bulge-to-total
  luminosity ratio (see \subsecn{decomposition}).  Columns (3) to (5)
  give average isophotal parameters $a_{3}/a$, $a_{4}/a$, and ellipticity
  in four different regions
  for each galaxy. Average values of isophotal parameters in Regions
  1, 2, 3 and 4 are listed, when available, in the first, second,
  third and fourth rows respectively for each galaxy.  }
\end{deluxetable}

Our method of obtaining characteristic parameter values is similar to the method used by
\citet{hao06} and \citet{ben88,ben89}, except that (i) \citet{hao06} 
considered a  single characteristic parameter value in the region $2r_{s} - 1.5R_{50}$ 
while \citet{ben88,ben89} defined the value 
 in region $2r_{s} - 1.5r_{e}$, where $r_{e}$ is the de Vaucouleur half light radius, and 
 (ii) \citet{ben88,ben89} estimated the characteristic value of ellipticity 
as the maximum value in a peaked ellipticity profile, and as the
value of the ellipticity at $1.5r_{e}$ in case of a continuously increasing or decreasing 
ellipticity profile. We 
have chosen multiples of  $R_{50}$ for defining the regions,
as the value of $r_{e}$ is very sensitive to the goodness of fit and the range of radius 
chosen for fitting surface brightness profiles 
with de Vaucouleurs law. The value of $R_{50}$ has been obtained from the SDSS DR7 galaxy 
catalogue given  by \citet{bla05}. 

\section{Observed properties of the galaxies}
\labsecn{observed properties}

\subsection{Some representative galaxies}
\labsubsecn{typical gals}

In \fig{ellipse over isophote} we show examples of three typical galaxies in our
sample. The left panels show the observed isophotes (in blue) in each region of the 
galaxy along the semi-major axis, superimposed on
grey-scale images from the LFC. The smooth red contours are the
best-fitted ellipses to the observed contours (see \secn{parameter estimation}). 
The right panels show the variation of $a_{4}/a$ along 
the semi-major axis of the galaxies. 

\begin{figure}
        \centering

\fbox{
\includegraphics[width=0.36\textwidth,angle=0.0]{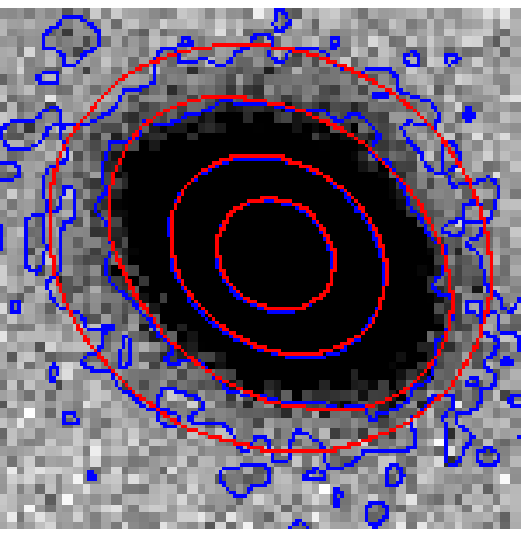}}%
\fbox{
\includegraphics[width=0.36\textwidth,angle=0.0]{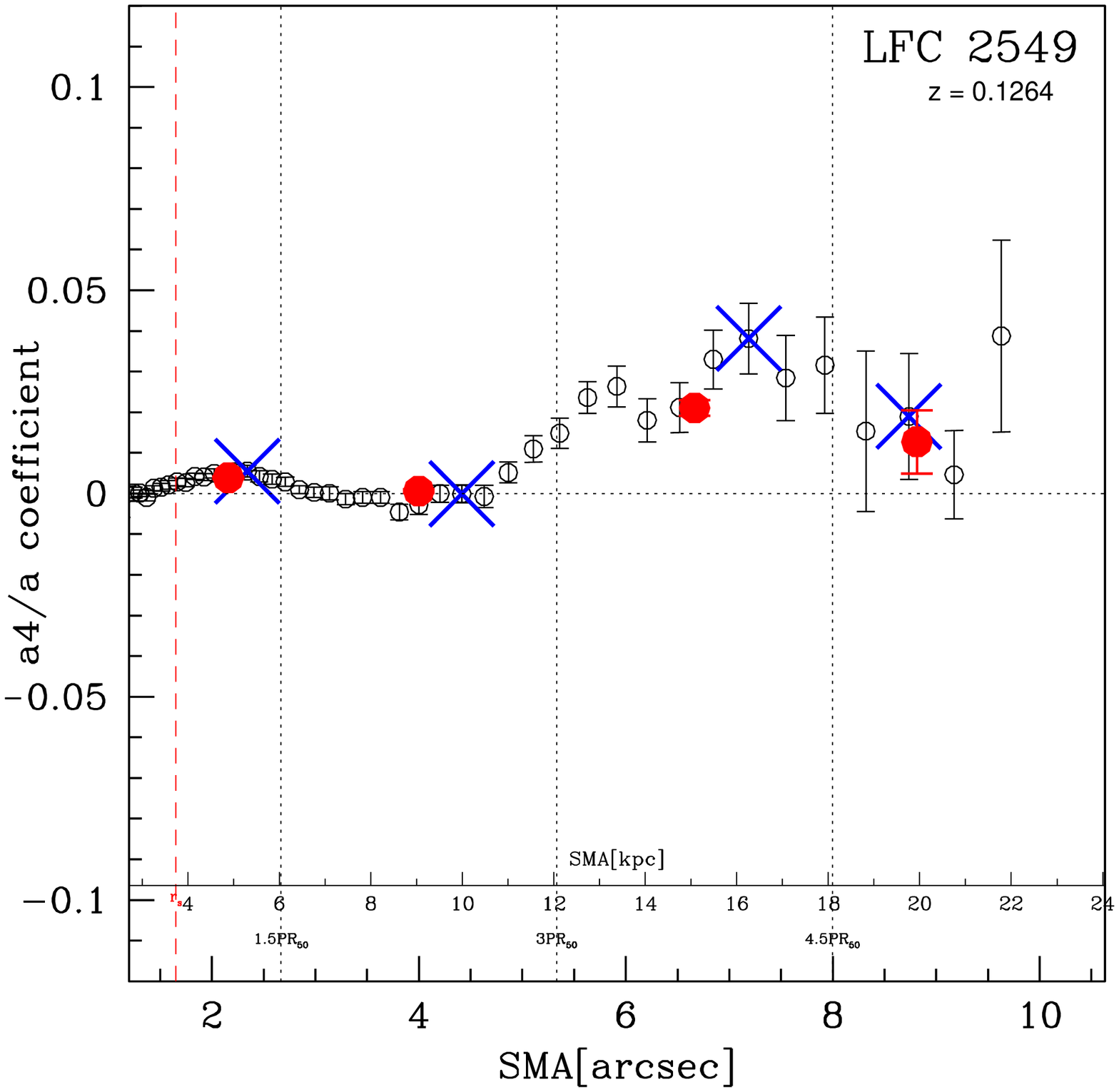}}%

\fbox{
\includegraphics[width=0.36\textwidth,angle=0.0]{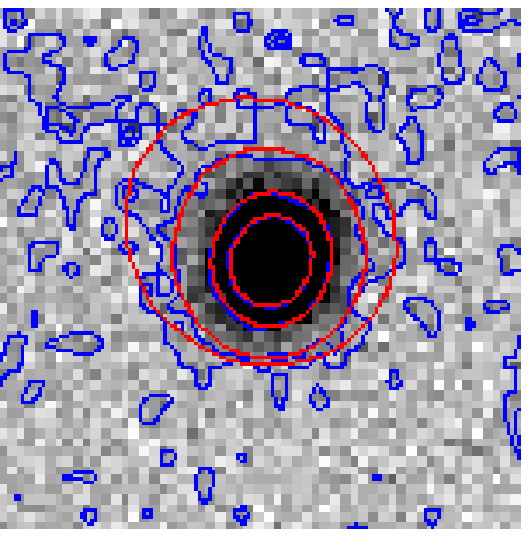}}%
\fbox{
\includegraphics[width=0.36\textwidth,angle=0.0]{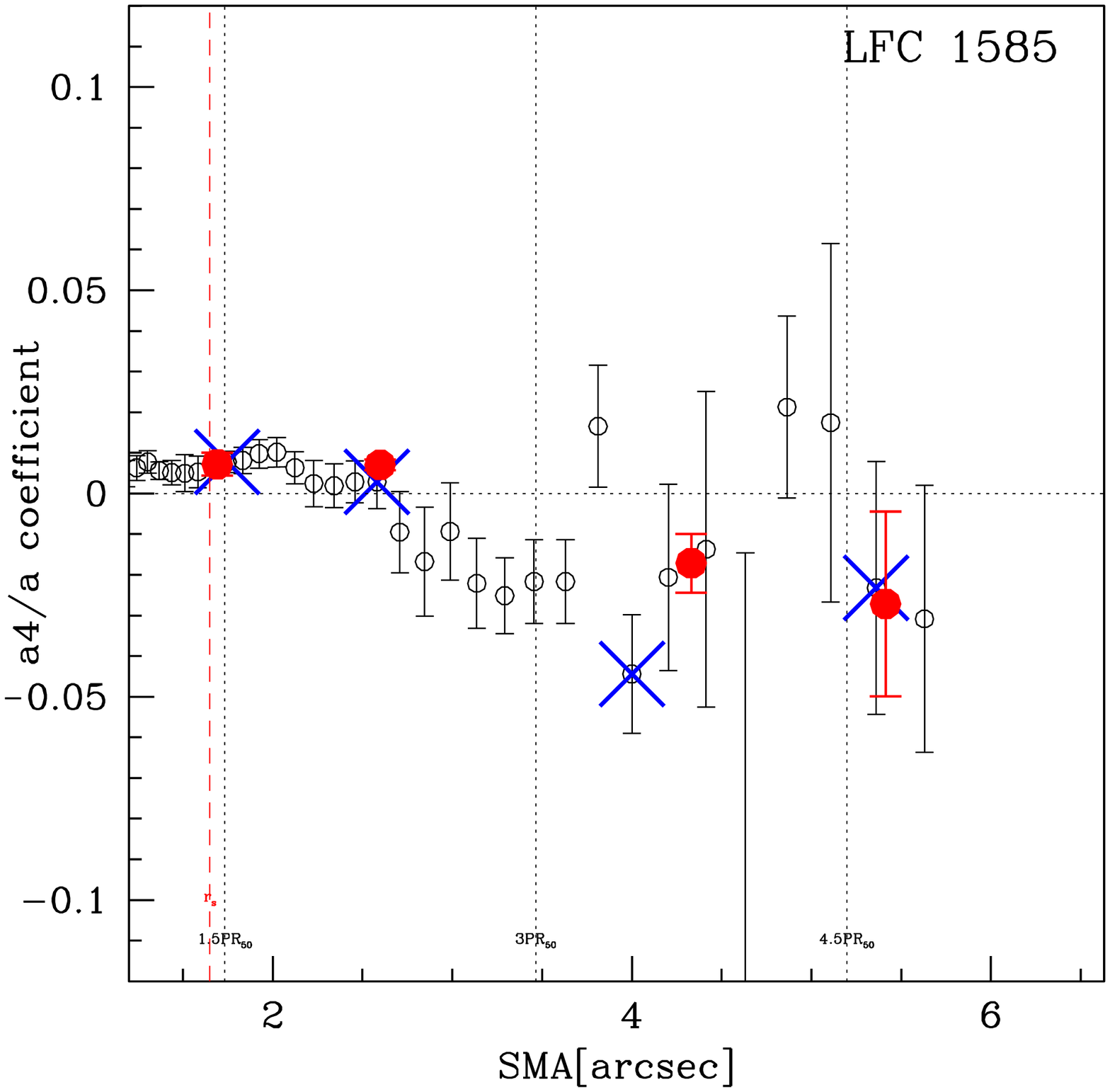}}%

\fbox{
\includegraphics[width=0.36\textwidth,angle=0.0]{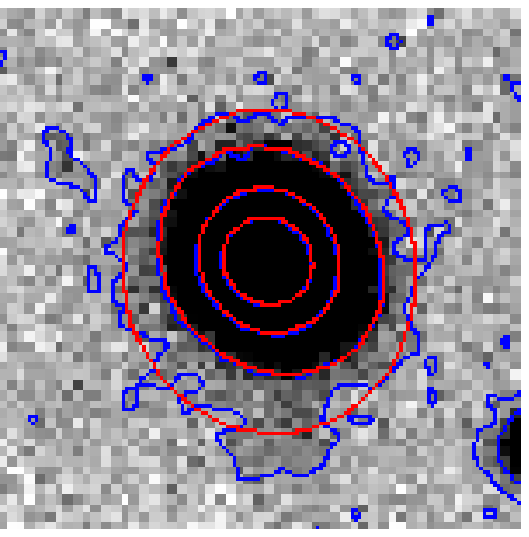}}%
\fbox{
\includegraphics[width=0.36\textwidth,angle=0.0]{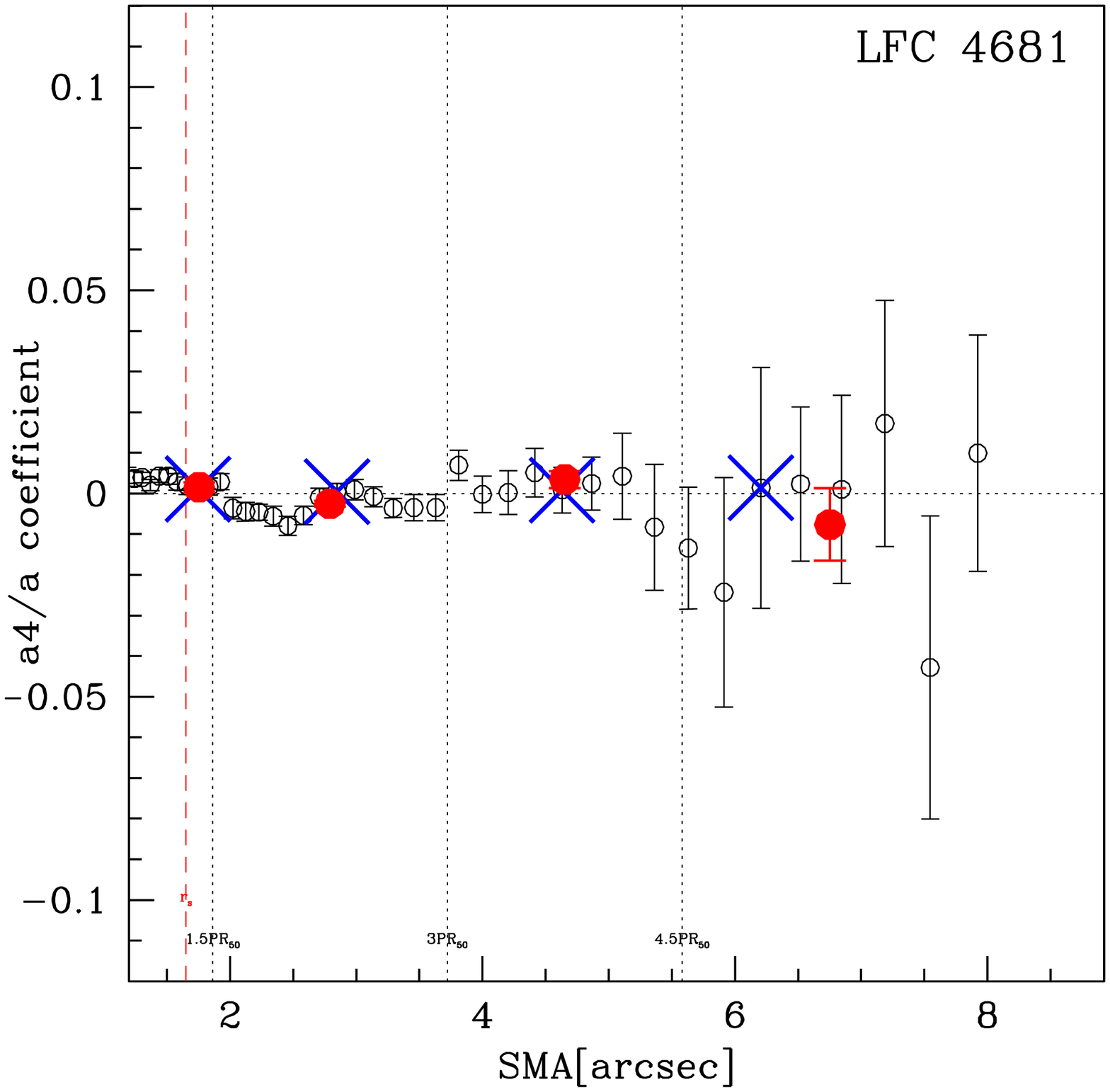}}

\caption{ {\small The left panels show {\it i}-band image cut outs
  ($20\arcsec \times 20\arcsec$) for three of our
  galaxies. Isophotes (in blue) are shown together with the
  best-fitted ellipses (in red). The right panels show the variation of $a_{4}/a$ along the
semi-major axis of each galaxy (with $1 \sigma $ error bars).
The  dotted vertical lines in black colour indicate the bins
of semi-major axis length in which the mean values of isophotal 
parameters are computed. 
A red circle shows the mean value of  $a_{4}/a$ in each bin. 
The positions of the blue contours in each left-hand panel are marked 
by blue crosses in the corresponding right-hand panels.
The region of each galaxy within the 
seeing radius (dashed vertical line)
is excluded from the study. Most previous studies of isophotal parameters 
have been confined to only the innermost of our four bins.} 
} 
\labfig{ellipse over isophote}
\end{figure}

These three examples represent common patterns for galaxies in our sample.
The $a_{4}/a$ profile for the galaxy LFC1208\_2549 
shows diskiness (positive $a_{4}/a$) in all the regions except in Region 2. 
The diskiness of galaxy LFC1208\_2549 peaks in Region 3. The profile 
of LFC1208\_1585 shows boxiness (negative $a_{4}/a$) 
in outer regions and small diskiness in the inner regions.
The profile for the  galaxy LFC1208\_4681 remains 
constant along the semi-major axis with value 
of $a_{4}/a$ close to zero, as the isophotes of this galaxy do not deviate much from 
the elliptical shape. The  vertical lines in black colour show the bins 
in which the mean values of isophotal parameters are computed. A red circle 
shows the mean value of  $a_{4}/a$ in each bin. The red dotted vertical line in each plot marks 
the seeing radius of the image frame. The region of each galaxy within the seeing radius is excluded from 
the study. The necessity of considering radial 
variation of isophotal parameters, as we do in this paper, is clear from the profiles 
in the figure.

\subsection{Sample properties}
\labsubsecn{sample properties}

In this section we describe some of the basic properties of our sample 
and compare those with the properties of other samples used for
similar studies. We look for evidence of systematic structural changes
between the inner and outer regions of galaxies.

We first examine the differential probability distribution of ellipticity
 $\epsilon (= 1 - b/a$), and shape parameters $a_{4}/a$ and $a_{3}/a$
in different regions of our sample galaxies.\\

We have fitted empirical Gaussian distribution functions (\equn{gaussian}) to the histograms of
various shape parameters for our galaxies, in different radial
regions. Although the intrinsic distributions are unknown {\it a
  priori}, empirical fits to the data may be useful for comparison 
between datasets or with theoretical studies.  
The results of the fits are given in \tablem{Gaussian} and 
the fits along with the distribution of the above parameters 
are shown in \fig{histogram ellip}. In some cases we
could get better fits by fitting multiple Gaussians but the
statistical significance and physical meaning of such fits was not
clear. The differential probability distribution we use 
for $\epsilon$, $a_{4}/a$ and $a_{3}/a$ are of the form
\begin{eqnarray}
p(x)dx = [ k_{1} + k_{2}x + k_{3} e^{-\frac{1}{2}(\frac{x-\mu_{1}}{\sigma_{1}})^{2}} ] dx \nonumber  \\
p(x)dx = [ k_{1} + k_{2}x + k_{3} e^{-\frac{1}{2}(\frac{x-\mu_{1}}{\sigma_{1}})^{2}}
             + k_{4} e^{-\frac{1}{2}(\frac{x-\mu_{2}}{\sigma_{2}})^{2}} ] dx,
\labequn{gaussian}
\end{eqnarray}
The coefficients $k_{3}$, $k_{4}$, $\mu_{1}$, $\mu_{2}$, $\sigma_{1}$ and $\sigma_{2}$ are 
the amplitude, mean and  the standard deviation of the fitted Gaussian. 
The coefficients $k_{1}$ and $k_{2}$ are zero point and slope of the baseline for the Gaussian.
The results of the fit are given in the  \tablem{Gaussian} and 
the fits along with the distribution of the above parameters 
are shown in \mfig{histogram ellip}{histogram a3}.

\begin{deluxetable}{lllllllllll}
\tablecaption{Values of coefficients for fitted single and/or double Gaussians \labtablem{Gaussian}
 (see \equn{gaussian} and \mfig{histogram ellip}{histogram a3})}

\tablewidth{0pt}
\tabletypesize{\scriptsize}

\tablehead{
\colhead{Parameter }& \colhead{$k_{1}$}& \colhead{$k_{2}$}& 
\colhead{$k_{3}$}&\colhead{$\mu_{1}$}& \colhead{$\sigma_{1}$}&  
\colhead{$k_{4}$}&\colhead{$\mu_{2}$}& \colhead{$\sigma_{2}$}& 
\colhead{$\chi^{2}$}&\colhead{$rms$}
} 
\startdata
$\epsilon$ (Region 1)  &
0.072     & -0.091  & 
0.158     &  0.111  &  0.057  & 
-     &  -  &  -   &
 1.05      & 0.025  \\
%
%
&
$\pm$0.056    & $\pm$0.131 &
$\pm$0.049     & $\pm$0.012  & $\pm$0.037   &  
    &   &   &
    &        \\ \hline  

$\epsilon$ (Region 2) &
0.036      & -0.041  & 
0.240      &  0.127  & 0.053  & 
-   & - & -&
1.45   & 0.023  \\
%
%
&
$\pm$0.045    & $\pm$0.094  &
$\pm$0.040    & $\pm$0.009  & $\pm$0.013 &  
              &             &            &
    &        \\ \hline

$\epsilon$ (Region 3)&
0.011      & 0.0 & 
0.202    & 0.159  & 0.083  & 
-   & - & - &
1.03   & 0.022    \\
%
%
&
$\pm$0.014    & $\pm$0.0  &
$\pm$0.017    & $\pm$0.008  & $\pm$0.012  &  
   &  &  &
    &        \\ \hline

$\epsilon$ (Region 4)  &
0.036     &   0.0  & 
0.210     &   0.160  & 0.051  & 
-   & - & - &
1.01   & 0.033  \\
%
%
&
$\pm$0.018    & $\pm$0.0  &
$\pm$0.032    & $\pm$0.009 & $\pm$0.020 &  
   &  & &
    &        \\ \hline \hline

$\frac{a_4}{a} $ (Region 1)&
0.0   &  0.0 & 
0.0486   & 0.0010 & 0.0131 & 
0.3143    & 0.0022   &  0.0042  &
 1.04  &  0.010   \\
%
%
&
$\pm$ 0.0  & $\pm$ 0.0 &
$\pm$0.0152  & $\pm$ 0.0098 & $\pm$ 0.0053  &  
$\pm$ 0.0205  & $\pm$0.0001 & $\pm$ 0.0005&
    &        \\ \hline

$\frac{a_4}{a} $ (Region 2)  &
0.0064    & -0.0726  & 
0.5163    & 0.0003  & 0.0016 & 
0.1256  & 0.0015  & 0.0093  &
1.14    & 0.013  \\
%
%
&
$\pm$0.3322    & $\pm$2.1186 &
$\pm$0.09308   & $\pm$0.0007 & $\pm$0.0024  &  
$\pm$0.5300    & $\pm$0.2297 & $\pm$0.2068 &
    &        \\ \hline

$\frac{a_4}{a} $ (Region 3)  &
  0.0149    &  -0.0273  & 
0.0683    & -0.0270 & 0.0011 & 
0.1240   & 0.00001  & 0.0079 &
0.92   & 0.020  \\
%
%
&
$\pm$0.0040   & $\pm$0.0300 &
$\pm$0.0632    & $\pm$0.0038 & $\pm$0.0002   &  
$\pm$0.0107    & $\pm$0.00002  & $\pm$0.0013 &
    &        \\ \hline

$\frac{a_4}{a} $ (Region 4) &
0.0268     & -0.0267 & 
0.1141    & -0.0256  & 0.0014  &
0.1407    &  0.0110  & 0.0030  &
0.53    &  0.021  \\
%
%
&
$\pm$0.0076    & $\pm$0.1084 &
$\pm$0.0810    & $\pm$0.0321 & $\pm$0.0005   &
$\pm$2.6219    & $\pm$0.0004 & $\pm$0.0010   & 
    &        \\ \hline   \hline
$\frac{a_3}{a} $ (Region 1) &
0.0      &  0.0 & 
0.4510      & -0.0009  & 0.0078  & 
0.0135 
  & -0.0112  & 0.0712  &
0.92    & 0.008     \\
%
%
&
$\pm$0.0   & $\pm$0.0 &
$\pm$0.0153 & $\pm$0.0003  & $\pm$0.0004 &  
$\pm$0.0068      & $\pm$0.0234   & $\pm$0.0208  &
    &        \\ \hline

 $\frac{a_3}{a} $ (Region 2)&
0.0      & 0.0  & 
0.3639      & -0.0011   & 0.0105  & 
-   & - & - &
1.38  & 0.023  \\
%
%
&
$\pm$0.0   & $\pm$0.0 &
$\pm$0.0307  & $\pm$0.0010  & $\pm$ 0.0009  &  
   &    &   &
    &        \\ \hline

$\frac{a_3}{a} $ (Region 3) &
0.0      & 0.0  & 
0.2253      & -0.0024     & 0.0142   & 
0.0593  & 0.0513   & 0.0140  &
1.01  & 0.012   \\
%
%
&
$\pm$0.0   & $\pm$0.0   &
$\pm$0.0112     & $\pm$0.0008     & $\pm$0.0009 &  
$\pm$0.0322    & $\pm$0.0043   & $\pm$0.0056   &
    &        \\ \hline

$\frac{a_3}{a} $ (Region 4) &
0.0    &  0.0 & 
0.1081    & 0.0046   & 0.0352    & 
0.0090   & 0.0291     & 0.0036    &
  0.93   & 0.023    \\
%
%
&
$\pm$0.0     & $\pm$0.0 &
$\pm$0.0131  & $\pm$0.0061      & $\pm$0.0062    &     
$\pm$0.0231      & $\pm$0.0302   & $\pm$0.0009     &
    &        \\ \hline 



\enddata
\end{deluxetable}

\begin{figure}[t]
        \centering
\includegraphics[width=1.0\textwidth,angle=0.0]{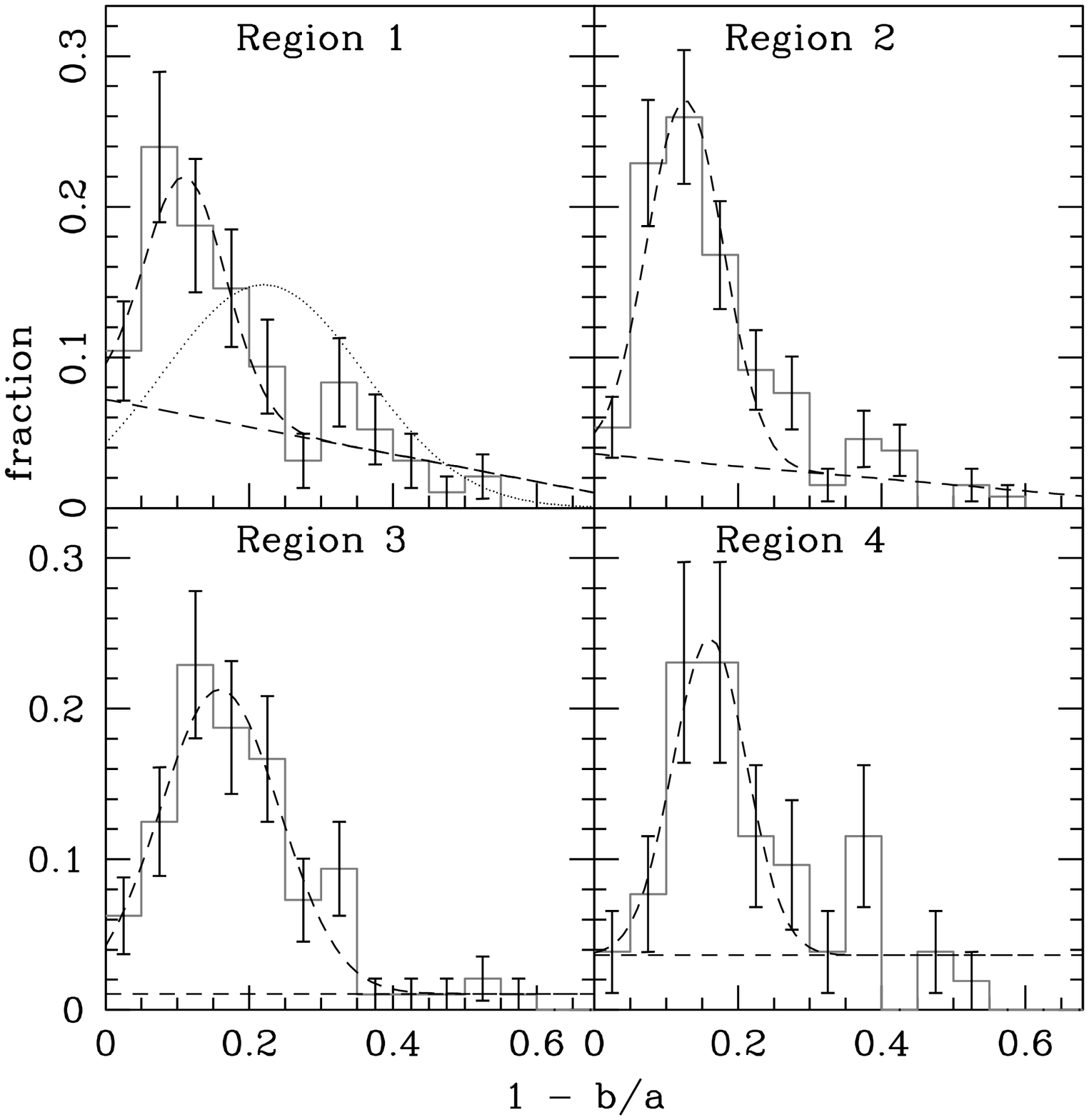}
\caption{Distribution of ellipticity $\epsilon = (1 - b/a)$, in different
  regions. The dashed lines are the best fitted Gaussian distributions with baseline
  (see text and \tablem{Gaussian}). The dotted curve in
  Region 1 is the best fit Gaussian to the ellipticity distribution obtained by
  \citet{hao06} for their sample. The error bars indicate $1\sigma$ Poisson errors. }
\labfig{histogram ellip}
\end{figure}

The ellipticity distribution of the sample used by \citet{hao06} (dotted
curve in Region 1, upper left panel of \fig{histogram ellip}) shows a
peak around $\epsilon\sim0.2$. The main peak in our sample, at
$\epsilon \sim 0.1$, indicates the presence of a larger fraction of
rounder galaxies. Both distributions of ellipticity drop to zero for
$\epsilon > 0.5$.  This difference in the ellipticity distributions is
presumably because of the very different data sets.  The galaxies of
\citet{hao06} are all large and relatively nearby, with $z < 0.05$,
selected over a large area of the sky. Most of our galaxies are much
smaller and more distant, with a redshift distribution peaking at $z
\sim0.1$ and extending to beyond 0.5 (see \fig{redshift}).

If Gaussian functions are fitted to the ellipticity distributions in
\fig{histogram ellip}, we find that as we go from Region 2 to Region 4, the
peak of the distribution occurs at around $0.13$, $0.16$ and $0.16$,
i.e. the peak shifts slightly towards more flattened
ellipses. Overall, the distributions of ellipticity in Regions 1 and 2
are similar to each other, as are those for Regions 3 and 4, but there
are significant differences between the inner and outer regions.

\fig{histogram a4} shows the distribution of the quadrupole parameter
$a_{4}/a$ for our sample.  The distribution in Region 1 shows a slight
excess of disky isophotes (positive $a_{4}/a$ values), although not as 
strong as that found at similar radii by \citet{hao06} and shown by the 
blue curve. However, the distribution we find in Region 2 (upper right panel) 
does resemble that of \citet{hao06}. As in the case of the ellipticity plots in \fig{histogram
  ellip}, the histograms for $a_{4}/a$ in Regions 1 and 2 are similar
to each other but quite different from those for Regions 3 and 4.

\begin{figure}
        \centering
\includegraphics[width=1.0\textwidth,angle=0.0]{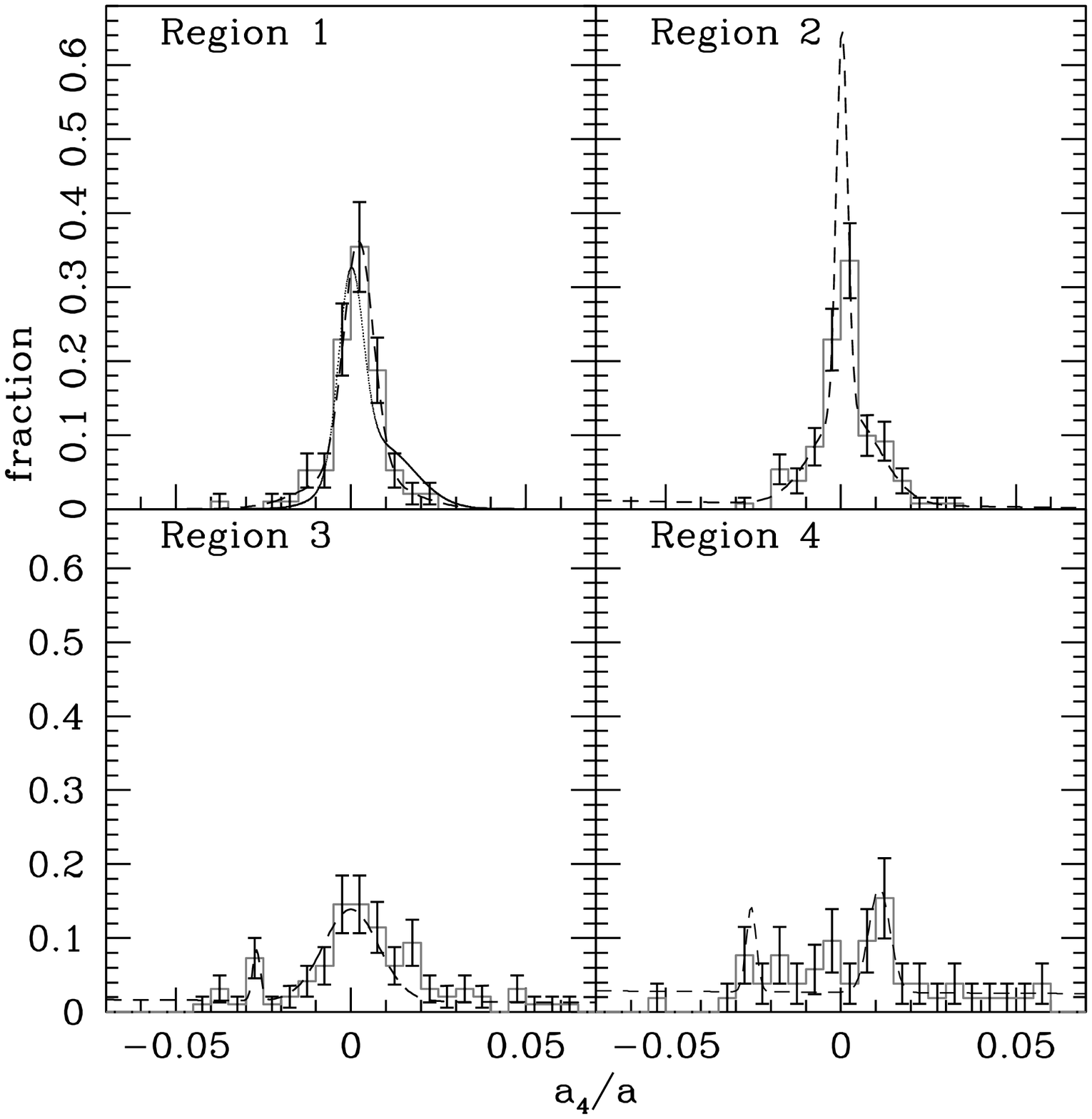}
\caption{ Distribution of $a_{4}/a$ parameters, in different
  regions. The dashed lines are best fitted single or double Gaussians,
  (see text and \tablem{derived}). The dotted curve in Region 1 is the best fit
  Gaussian to the $a_{4}/a$ distribution obtained by
  \citet{hao06} for their sample. The error bars indicate $1\sigma$ Poisson
  errors.
}
\labfig{histogram a4}
\end{figure}

The $a_{3}/a$ parameter quantifies the deviation from pure ellipse that occur along the 
observed isophote every $120\degr$. \citet{pas07} 
have suggested that such deviations may be attributed to the 
presence of dust or clumps present within the galaxy. But the distribution 
of the dust can be irregular in the galaxy and can also give rise to higher order
Fourier coefficients. It may be better to interpret $a_{3}/a$ as a structural 
parameter, possibly indicating dynamical effects from galaxy interactions or mergers.
\fig{histogram a3} shows 
the distribution of this parameter for our sample.

The distribution of $a_{3}/a$ in Region 1 is reasonably fitted by a 
sum of two Gaussian functions with mean $-0.0009 $ and $-0.0112 $ with dispersions
$0.0078$ and $0.0712$. The value of $\chi^{2}$ per degree of freedom for the fit
is $\chi^{2}_{\nu}=0.92$. In comparison, the distribution of the 
 \citet{hao06} sample is reasonably fitted by a single Gaussian with zero mean 
and dispersion of 0.0032. In our sample the
peaks of distribution in Region 1 and Region 2 occur at negative values of the
parameter $a_{3}/a$ while one peak of the double Gaussian in Region 3 and both
peaks of the double Gaussian in Region 4 occur at positive values of 
parameter $a_{3}/a$ (see \tablem{Gaussian}). 
It is not known as to which property of galaxy determines the sign of $a_{3}/a$ parameter. 
The findings of \citet{jog06} suggest that
non zero values of $a_{3}/a$ parameter along with the presence of lopsidedness 
in the inner regions of galaxies can be interpreted as a signature of
dynamically unrelaxed behaviour in the inner regions of the galaxies. These authors
have Fourier-analysed the central few kpc of advanced mergers
of galaxies using images from the Two-Micron All-Sky Survey (2MASS) and have obtained 
amplitudes and phases of Fourier components m = 1, 2, 3 and 4. 
Their analysis indicates that in the case of mergers, the 
amplitudes A1 and A2 (for m=1 and 2 respectively) dominate over m=3 and 4, and A3 is 
important only when A1 is large. A1 denotes the amplitude for the 
lopsidedness and is an indicator of mass asymmetry measured with respect to the constant 
center (see Section 3.2 of their paper for details). The study of  
lopsidedness in our sample galaxies should reveal whether the peaks 
of the $a_{3}/a$ distribution around non-zero values in the 
inner regions of our galaxies can be interpreted as a signature of 
dynamically unrelaxed inner regions.
Though we do not understand the reason for the 
occurrence of peaks at positive values of $a_{3}/a$  in Region 3 and in Region 4,  
we can say that the processes which affect isophotal shape are different in 
the inner and outer regions.

\begin{figure}
        \centering
\includegraphics[width=1.0\textwidth,angle=0.0]{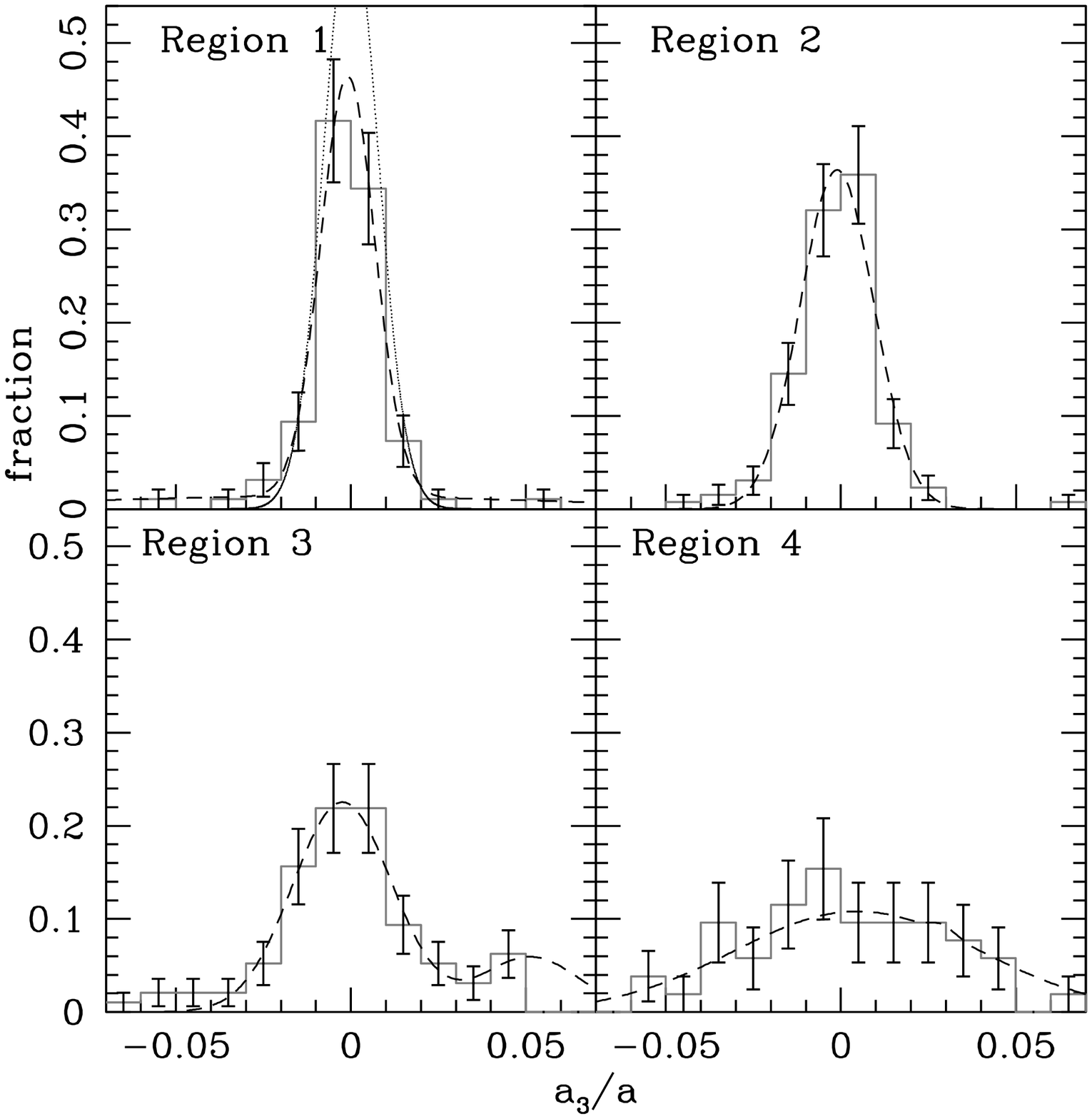}
\caption{Distribution of $a_{3}/a$ parameters, in different regions. The dashed lines 
are best fitted single or double Gaussians (see text and \tablem{Gaussian}). The 
dotted curve in Region 1 is the best fit Gaussian to
the $a_{3}/a$ distribution obtained by \citet{hao06} for their sample.
The error bars indicate $1\sigma$ Poisson errors.}
\labfig{histogram a3}
\end{figure}

 
We next look for correlations between the values for various
shape parameters for individual galaxies, when adjacent
radial regions are compared.

\fig{region ellip} shows that there is a strong correlation between
the ellipticity parameters in the innermost Regions 1 and 2 (upper
 panel). However, there is a significant change going from Region
2 to Region 3 (lower left); while most of the higher ellipticity
galaxies remain correlated (those with $\epsilon > 0.3$ in Regions 1
and 2), about half of the nearly circular galaxies become more
flattened. The outer most parts (Regions 3 and 4, lower right panel) show more
scatter and fewer galaxies have data in Region 4, but an overall
correlation remains. 

\begin{figure}
\centering
\includegraphics[width=1.0\textwidth,angle=0.0]{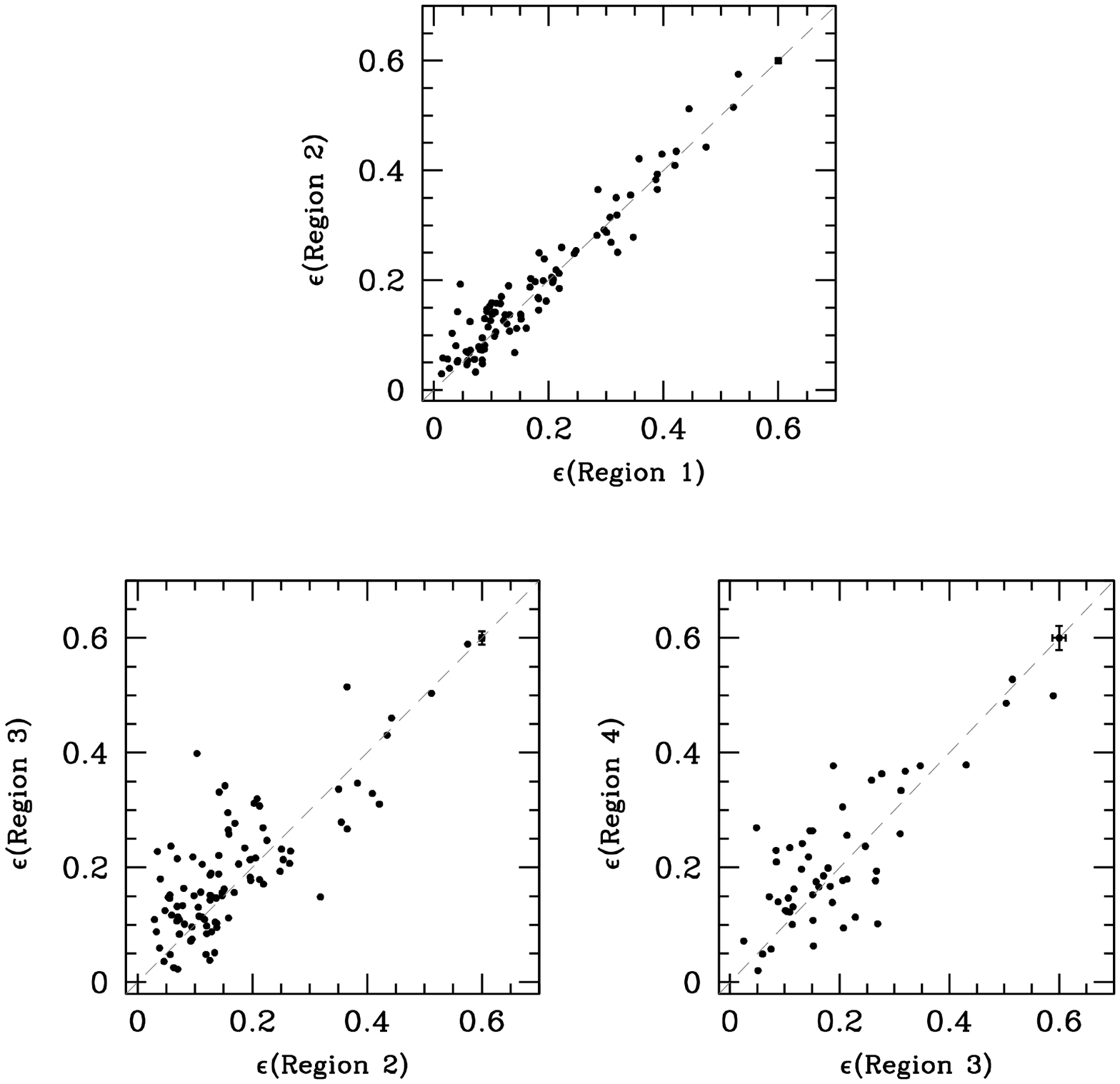}
\caption{Correlation of ellipticity ($\epsilon = 1 - b/a$) parameters from neighboring regions.
 The median error bars are shown at the the top right-hand corner in each pannel.
}
\labfig{region ellip}
\end{figure}

We show similar diagrams for the shape parameter $a_{4}/a$ in
\fig{region a4}. The dominant feature is a large increase in the
scatter of points between Regions 2 and 3, shown in the lower right
panel. The other two plots, for the outer Regions 3 and 4 (upper panel) 
and the inner Regions 1 and 2 (lower left) show some correlation between
adjacent regions, with few points populating the top left and bottom right quadrants.

\begin{figure}
\centering
\includegraphics[width=1.0\textwidth,angle=0.0]{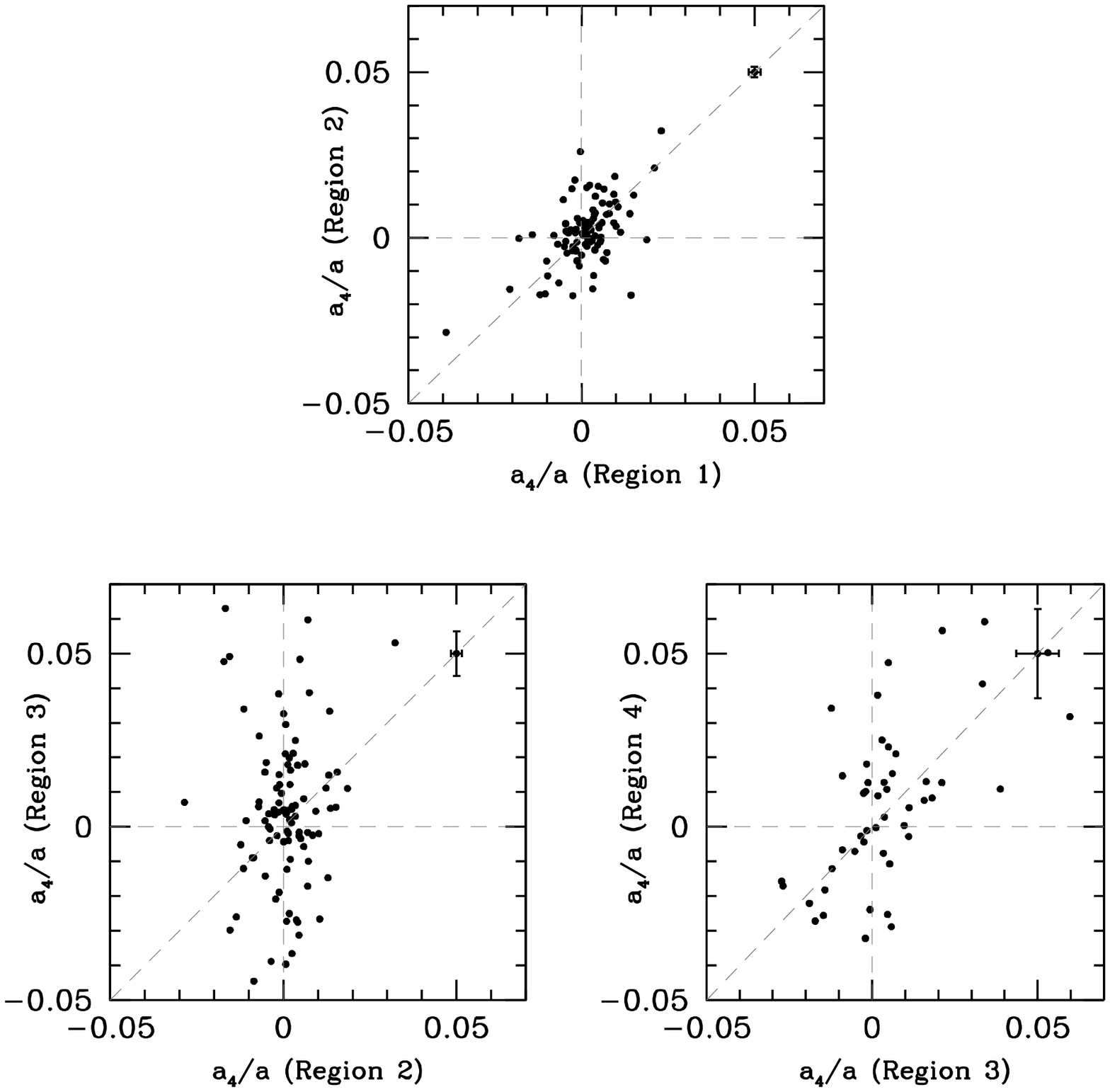}
\caption{The correlation of $a_{4}/a$ parameters from neighboring
  regions. The error bars are as in \fig{region ellip}.
}
\labfig{region a4}
\end{figure}

\subsection{Are distribution of isophotal parameters statistically different in different regions?}
\labsubsecn{parameters in regions}

\begin{deluxetable}{ c c c c c c c c c}
\tablecaption{Results of two sample K-S test for the $a_{4}/a$ parameter \labtablem{K-S test a4/a}}
\tablewidth{0pt}
\tablehead{
\multicolumn{1}{c}{Sample}& 
\multicolumn{2}{c}{ Region 1}& 
\multicolumn{2}{c}{ Region 2}& 
\multicolumn{2}{c}{ Region 3}& 
\multicolumn{2}{c}{ Region 4} 
} 
\startdata
    & $D$\tablenotemark{1}  & $P$\tablenotemark{2} & $D$ & $P$ & $D$ & $P$ & $D$ & $P$ \\   \hline
\citet{hao06} & 
0.15 & 0.041 & 

0.13 & 0.038 &

0.22 & 5.97e-04 &

0.29 & 5.81e-04 \\ \hline
Region 1 &
       &          &
0.08   & 0.855 &
0.25   & 0.005 &
0.35   & 6.41e-04 \\ \hline 
Region 2 &
       &         &
       &         &
0.21   & 0.014   &
0.33   & 5.44e-04  \\ \hline

Region 3    &
     &      &
     &      &
     &      &
0.16 & 0.383 \\ \hline

\enddata
\tablenotetext{1}{The K-S statistic $D$ is defined as the maximum value of the absolute difference between two cumulative 
distribution functions where the cumulative distribution function is obtained from the list of data points
of each sample on which K-S test is applied.}
\tablenotetext{2}{$P$ gives the level of significance with which the null hypothesis 
may be accepted. Small values of $P$ imply that the cumulative distribution function
of two samples tested are significantly different \citep[see][]{pre92}.
The confidence that both the population do not belong to the same parent distribution 
is given by $(1 - P) \times 100 $. 
}
\end{deluxetable}

\begin{deluxetable}{ c c c c c c c c c}
\tablecaption{Results of two sample K-S test for the $a_{3}/a$ parameter \labtablem{K-S test a3/a}\tablenotemark{1}}
\tablewidth{0pt}
\tablehead{
\multicolumn{1}{c}{Sample}& 
\multicolumn{2}{c}{ Region 1}& 
\multicolumn{2}{c}{ Region 2}& 
\multicolumn{2}{c}{ Region 3}& 
\multicolumn{2}{c}{ Region 4} 
} 
\startdata
    & $D$  & $P$ & $D$ & $P$ & $D$ & $P$ & $D$ & $P$ \\   \hline
\citet{hao06} & 
0.30 & 2.85e-07 &
0.30 & 1.65e-09 & 
0.33 & 1.16e-08 &
0.44 & 1.34e-08 \\ \hline
Region 1        &
     &          &
0.10 & 0.613    & 

0.21 & 0.031    &
0.29 & 0.007    \\ \hline
Region 2        &
     &          &
     &          &
0.19 & 0.034    &
0.28 & 0.005    \\ \hline
Region 3        &
     &          &
     &          &
     &          &
0.19 & 0.151    \\ \hline
\enddata
\tablenotetext{1}{Notation as in table \tablem{K-S test a4/a}.}
\end{deluxetable}
\begin{deluxetable}{ c c c c c c c c c}

\tablecaption{Results of two sample K-S test for the $ (1 - b/a) $ parameter \labtablem{K-S test ellipticity}\tablenotemark{1} }
\tablewidth{0pt}
\tablehead{
\multicolumn{1}{c}{Sample}& 
\multicolumn{2}{c}{ Region 1}& 
\multicolumn{2}{c}{ Region 2}& 
\multicolumn{2}{c}{ Region 3}& 
\multicolumn{2}{c}{ Region 4} 
} 
\startdata
    & $D$ & $P$ & $D$ & $P$ & $D$ & $P$ & $D$ & $P$ \\   \hline
\citet{hao06} &
0.28  & 3.31e-06&
0.28  & 3.93e-08 &
0.20  & 0.002    &
0.14 & 0.30    \\ \hline
Region 1        &
      &         &
0.11  &  0.466  &
0.19  &  0.068  &
0.27  &  0.002  \\ \hline
Region 2        &
      &         &
      &         &
0.16  & 0.098   &
0.23  & 0.039   \\ \hline 
Region 3        &
      &         &
      &         &
      &         &
0.13  &  0.574  \\ \hline 
\enddata
\tablenotetext{1}{Notation as in table \tablem{K-S test a4/a}.}
\end{deluxetable}
We performed a two-sample Kolmogorov-Smirnov test (K-S test) to 
investigate whether the distribution
of $a_{4}/a$, $a_{3}/a$ and $\epsilon$ are same in our four regions. (see
\dtablem{K-S test a4/a}{K-S test a3/a}{K-S test ellipticity}).

We used {AstroStat}\footnote{http://vo.iucaa.ernet.in/$\sim$voi/AstroStat.html},
 which uses a public-domain statistical computing package 
{\it R}\footnote{http://www.r-project.org/} for statistical analysis, to perform the K-S test,
in which the null hypothesis that the two samples belong to the
same parent distribution is tested. The strength of evidence in support of a null hypothesis 
is given by probability {\it P} in the table. 
If the {\it P} is less than the assumed {\it level of significance} the hypothesis is rejected, 
where the {\it level of significance} is the probability of making a decision
to reject the null hypothesis, when the null hypothesis is true. 

We find from the K-S test that the distribution of $\epsilon$ in  
neighboring Regions belongs to the same parent distribution, as the test is 
accepted at the $5\%$ level of significance. 
The distribution of $a_{4}/a$ and $a_{3}/a$ in 
Region 2 and Region 3 may not belong to the 
same parent distribution, as the test is rejected at the
$5\%$ level of significance. 
The distribution of $a_{4}/a$ and $a_{3}/a$ parameter in 
Region 1 and Region 2 are from same parent population,
 as suggested by high 
{\it P-value} (=0.86 and 0.61) obtained from the K-S test.
Similarly, the distribution of $a_{4}/a$ and $a_{3}/a$ parameter
in Region 3 and Region 4 are same as the test is accepted
at the $5\%$ level of significance, though the strength of evidence
in support of the null hypothesis is not as strong as it is 
for Region 1 and Region 2.
The Results of K-S test clearly shows the discontinuity in the 
distribution of parameters
as we go from Region 2 to Region 3.


\subsection{Frequency of boxy and disky ellipticals}
\labsubsecn{frequency of boxy}

\begin{deluxetable}{rlclll}
\tablecaption{Frequency of boxy and disky isophotes \labtablem{frequency}}
\tablewidth{0pt}

\tablehead{\colhead{Radial bin} & \colhead{Region} & \colhead{Number of galaxies} &  \colhead{boxy} & \colhead{disky}& 
\colhead{$\mid$ $a_{4}/a$ $\mid$ $\leq 0.2\%$} \\

\colhead{} & \colhead{} & \colhead{with data in region} &  \colhead{} & \colhead{}& 
\colhead{}
} 
\startdata
$r_{s}$ - 1.5$R_{50}$ &  Region 1 & 96 & 25 \%       & 48\%       & 27\%                    \\ \hline
1.5$R_{50}$ - 3.0$R_{50}$ & Region 2 & 131 & 28\%      & 43\%      & 29\%                   \\ \hline
3.0$R_{50}$ - 4.5$R_{50}$ & Region 3 & 96 & 55\%      & 36\%       & 09\%                     \\ \hline
r  $>$ 4.5$R_{50}$ & Region 4 & 52 & 54\%      & 40\%       & 06\%                   \\ \hline
\enddata
\end{deluxetable}

Bender et al. (1989) found that $\sim$ 1/3 of their sample galaxies show boxy isophotes, 
$\sim$ 1/3 pointed isophotes and $\sim$ 1/3 of isophotes have deviation smaller 
than 0.2\% of the semi-major axis length. The frequency  of
boxy and disky isophotes in different radial regions of our sample galaxies
is given in \tablem{frequency}. We find a larger 
fraction of disky isophotes in Regions 1 and 2. 
The diskiness is generally attributed to the presence of a weak edge-on disk superposed
on the spheroidal main body. One of the possible reasons for the increased fraction of 
disky isophotes in Region 1 is 
detection of those weak disks in our sample galaxies which would not be 
detected with relatively low SNR images. 

The major fraction of our galaxies have either boxy
or disky isophotes in their outer regions: we find that the 
fraction of near-circular galaxies having $\mid$ $a_{4}/a$ $\mid$ $\leq 0.2\%$
is less than $10\%$ in Regions 3 and 4, compared with $25\%$ to $30\%$ in 
Regions 1 and 2. This is consistent with 
the expectation that tidal extensions and other environmental effects
are likely to be stronger in the outer regions 
as compared to the inner regions of galaxies. 

We also find a higher frequency of boxy as compared 
to disky isophotes in the outer regions. It has been suggested by
\citet{nie89} that tidal extensions may also 
cause pointed isophotes. The higher fraction of 
boxy isophotes indicates that: (i) the presence of 
additional forces along with the tidal extensions; and (ii) that such
forces are either more frequent or more effective in outer regions.  

The frequency of disky isophotes is higher in the outer regions of dwarf early-type
galaxies having $\absm{B}> 17.0 $ in our sample. The fractions of disky isophotes 
for such galaxies in Region 1, Region 2,  Region 3 and 
Region 4 are $1/6$, $4/8$, $4/6$ and $2/2$ respectively.
The denominators of above fractions for each region are total number of 
dwarf early-type galaxies for which isophotal parameters are available.
This higher fraction of disky isophotes in outer regions from Region 1 to 
Region 4  observed for dwarf early-type galaxies in our sample
is quite different from the corresponding frequency in different regions  
when we consider all the galaxies in the sample.
Differences in the environment of the  dwarf and luminous early-type galaxies 
may explain the observed differences of their isophotal properties, as 
dwarf elliptical galaxies are found in high-density regions, either in galaxy clusters
or in locations near more massive spiral and elliptical galaxies \citep{fer89,van04}.

\subsection{Correlation between isophotal shape and ellipticity}
\labsubsecn{a4 and ellipticity}

\fig{Bender like a4_ellip} shows plots of $a_{4}/a$ against
ellipticity $\epsilon$ for galaxies in each of our four radial
regions.  Our data are most complete for Region 2; we miss some
galaxies in Region 1 when the 'seeing' $r_{s} > 1.5R_{50}$, and from
Regions 3 and 4 due to confusion and low S/N.
The solid lines in each panel define a chevron-shaped 
region within which all galaxies from the sample of \citet{ben89}
were found. The dotted lines are extrapolations of these lines to 
higher values of ellipticity.

\begin{figure}
        \centering
\includegraphics[width=1.0\textwidth,angle=0.0]{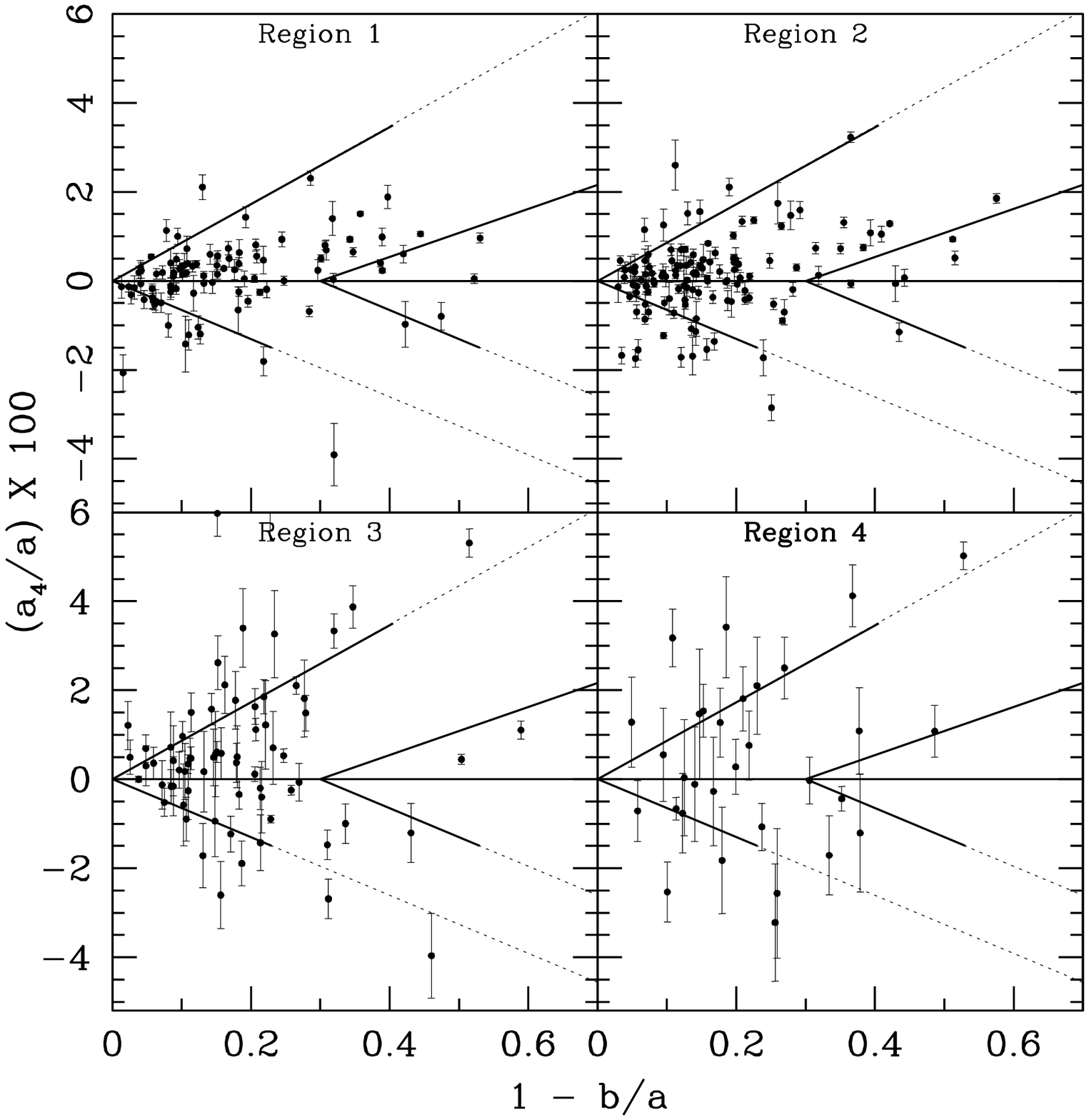}
\caption{Isophote shape parameter $a_{4}/a$ plotted against
  ellipticity, in the four radial regions. In Regions 3 and 4 we omit 
  galaxies with errors $\sigma(a4/a) > 0.01$ and $0.015$ respectively, for clarity.
\citet{ben89} found that their galaxies lay within the chevron-shaped
region between the pairs of solid diagonal lines. 
}
\labfig{Bender like a4_ellip}
\end{figure}

The distributions in Regions 1 and 2 are similar to each other, and
most galaxies lie within the zone defined by \citet{ben89}. Moreover,
our galaxies show the same trend, with a preponderance of disky
isophotes (with positive $a_{4}/a$) which extend to larger values as
the ellipticity increases. However, we do find some galaxies lying
outside \citet{ben89}'s boundaries, in particular a group with boxy
isophotes (negatative $a_{4}/a$) and low ellipticity in Region 2. We
also find some galaxies with large ellipticity ($1-b/a > 0.4$) but
with $a_{4}/a$ near zero.

Our data become increasingly incomplete and with larger errors in
Regions 3 and 4, where we have plotted only points with errors less
than 0.01 and 0.015 respectively.
The pattern appears similar in the two lower panels of \fig{Bender like
 a4_ellip}, and different from the pattern in the inner regions,
although this may be at least partly due to the increasing errors and smaller
samples. 

Overall, the isophote shape data are consistent with the patterns seen
in \subsecn{frequency of boxy} and clearly indicate that
the phenomena  
responsible for keeping the isophotes close to an elliptical shape 
at small radii (of the rounder galaxies) weaken as 
we go further out along the radius. Another possibility is that the effects 
of tidal interaction become more dominant, to make the isophotes deviate 
from the elliptical shape. Environmental studies may show whether this is true.

\section{Discussion}
\labsecn{discussion}
\subsection{Are projection effects important?}
\labsubsecn{projection effects}
The results obtained by  \citet{ben89} indicate
beyond  doubt that the boxiness or diskiness is an intrinsic property of the 
early-type galaxies rather than  an effect caused by projection. 
This means that the sign of the $a_{4}/a$ parameter is independent 
of projection effects, but it is not possible to say how the absolute value of $a_{4}/a$ depends
on the viewing angle. The trend found by \citet{ben89} and \citet{hao06}, 
that the larger values of $|a_{4}/a|$ are observed for galaxies which appear 
more elliptical is consistent, with a few exceptions, with the results obtained for 
our sample galaxies in \subsecn{a4 and ellipticity}.

\subsection{Comparison with N-body simulations}
\labsubsecn{N-body models}

The differences in the properties of boxy and disky 
elliptical are also supported by the N-body simulations of major mergers
\citep[e.g.][]{her93,lim95,kho05}, but some of the findings of N-body simulations
are not consistent with the observations. For example, \citet{sti91} and \citet{hey94} analyzed 
the isophotal shapes of the remnants from their N-body simulations of 
dissipationless collapse, and found that the same object 
has value of $a_{4}$ negative or positive depending on the viewing angle. 
Similarly, \citet{fab93} have reported that isophotal shape can not
discriminate between the possible origin of the early-type galaxies,
at least in the range $-1 \leq a_{4}/a \times 100 \leq 1 $ because
of the dependence of the shape upon the viewing angle as seen using
simulations. We do not understand the dependence of  $a_{4}/a$
parameter on viewing angle found by above authors which is inconsistent
with the observations. In future, the understanding of true intrinsic shapes 
of early-type galaxies may elucidate the effect of projection on the 
value of $a_{4}/a$ parameter.

N-body merger simulations have been widely used by various researchers to study the 
origin of boxy and disky ellipticals. Simulations have been done taking
progenitors for the merger as (i) a combination of galaxies with different 
morphologies (ii) a combination of galaxies with different mass ratios.
\citet{bou05} have shown in their Fig. 2,  that a 7:1 merger produces a  
galaxy with boxy-isophotes in the inner region while the outer 
region is highly disky.
Such radial variation has also been reported using the analysis of 
2MASS data for the Arp mergers \citep[see][Appendix]{chi02}.
Similar merger scenarios can be considered as possible origin
of the inner boxy and outer disky isophotes observed in some of 
our sample galaxies.

\citet{bou05} have shown that for a merger with 7:1 mass ratio,
outer diskiness is observed for a $25 \magarcsecsqi$ isophote 
around $20 \kpc$ from the centre (see their Fig. 2 \& 3).
In the region $\sim 20 - 30 \kpc$, the surface density falls by a 
factor of $\sim 80 - 100$ and for a constant mass-to-light ratio this 
corresponds to a $\sim 4 - 5$ magnitude difference. The outer region of 
\citet{bou05} corresponds 
to the Region 4 of most of our sample galaxies, thus a disk is clearly 
indicated at such large radii.

A typical value of $a_{4}/a \simeq 0.01 - 0.02$ is obtained for the remnants 
of collisionless N-body simulations of binary mergers of disk galaxies 
with mass ratios of 1:1, 2:1, 3:1 and 4:1 \citep[see][Fig. 3]{naa03},
while observations show larger diskiness in early-type galaxies. For example, 
observed values $a_{4}/a \la 0.03$ are obtained for the galaxies studied by \citet{ben89},
while some of our sample galaxies show larger diskiness ($a_{4}/a > 0.03$). Such 
galaxies are also a part of sample studied by the \citet[][Fig. 1]{hao06b}.
The origin of strong diskiness therefore should have a different merger scenario from that 
studied by \citet{naa03}. The galaxies with larger diskiness could well be explained 
by hybrid mergers, in mass ratio 4:1-10:1 \citep{bou04,bou05}. They report
$a_{4}/a = 0.064 \pm 0.01$ for mergers with mass ratio 4.5:1 and $a_{4}/a \simeq 0.07$
for mergers with mass ratio 7:1.

In binary merger scenarios, disky ellipticals are produced by 3:1-4:1
mergers while boxy ellipticals are product of equal-mass 1:1 mergers
\citep{bou07}. But equal-mass binary mergers fail to reproduce
the most boxy elliptical galaxies, particularly giant ellipticals
\citep{naa03,naa09}. Alternatively, multiple minor mergers can reproduce 
boxy ellipticals, where the product of mergers mainly depends
on the total merged mass. When total merged mass increases ($\geq 1.6$),
on average the remnants tend to show boxy isophotes \citep[see][Fig. 5]{bou07}.
The merged mass has been defined in a particular way by authors in their 
paper: for instance, when the initial galaxy has merged with 3 companions,
each of them having a 5:1 mass ratio, this so-called merged mass is
1.6 (1.0 for the main initial galaxy and 0.2 for each companion).
\fig{a4 vs i mag}
shows a plot of $a_{4}/a$ with absolute {\it i}-band magnitude
for the galaxies with redshift information. We find that a larger number 
of galaxies with boxiness have $\absm{i}< -19.5$, 
in all regions along the radius. 
Since more luminous galaxies will be more massive, the above trend 
can be considered similar to the trend observed for boxiness of the 
remnants of multiple minor mergers observed by \citet{bou07}, though 
we do not exactly know the mass corresponding to $\absm{i}<-19.5$.


\begin{figure}
        \centering
\includegraphics[width=1.0\textwidth,angle=0.0]{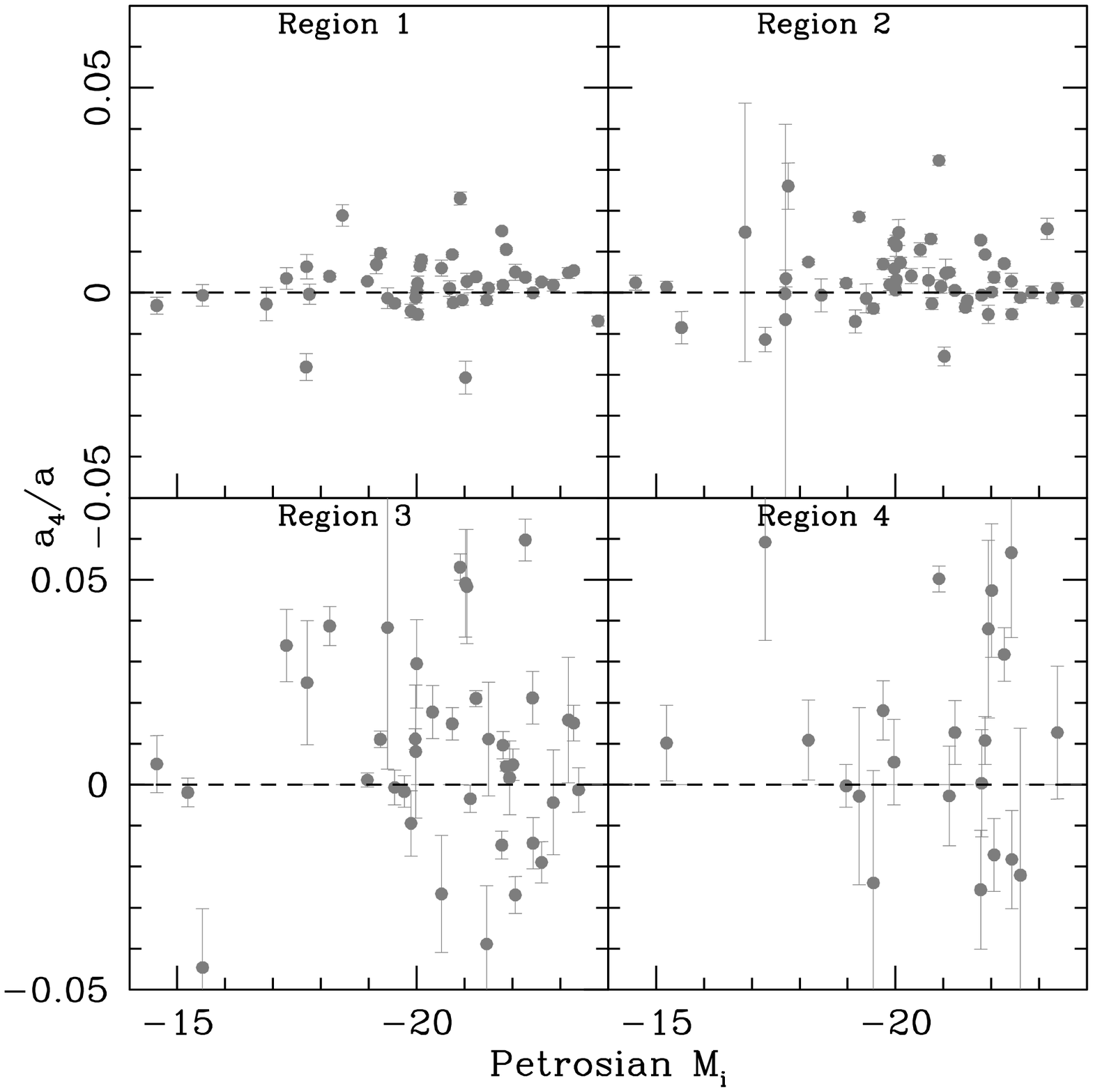}
\caption{Isophotal shape parameter $a_{4}/a$ plotted against absolute Petrosian i magnitude.}
\labfig{a4 vs i mag}
\end{figure}

We have tried to compare our observations with the 
isophotal shapes of the merger remnants, but it is not in the scope 
of this paper to trace the merger history of the sample galaxies
on the basis of observed isophotal shapes.

\section{Conclusions}
\labsecn{conclusion}
In this paper, we have studied the properties of the isophotal shapes of
early-type galaxies (E/SO) to very faint outer regions, well beyond the
levels reached by previous studies of this type. Our study is based on
deep 5-color Hale Telescope LFC CCD images of several fields, originally
taken for a different project, that enable us to reach surface
brightnesses some $4 \magarcsecsqi$ fainter than most similar studies. Here
we present the first results for one $14\arcmin \times 25\arcmin$  field.

We selected target galaxies using the deep LFC {\it i}-band image of the field and
used much shallower SDSS imaging data to get basic parameters such as apparent
magnitudes and Petrosian radii for each galaxy. 
266 sufficiently large and bright galaxies were selected
for further study, 132 of which were identified as being of early-type on
the basis of their bulge-to-total light ratios. We obtained spectra for
over half of this sample using the multi-fibre system AAOmega on the AAT.
These yielded reliable redshifts for 53 of the early-type galaxies,
enabling us to derive their absolute magnitudes and physical sizes.

We fitted a sequence of ellipses to successive isophotes in the deep LFC
images and derived a range of isophotal shape parameters that measure
their ellipticity and orientation, and also higher order departures from a
purely elliptical shape. We then derive mean values for these parameters
in four radial bins along the semi major axis of each galaxy. We find
empirical fitting formulae for the probability distribution of the
different isophotal parameters in each bin, which will be useful for
comparison with theoretical studies, e.g. from N-body simulations.

Finally, we have investigated possible correlations of isophotal shape
parameters with other global properties of the galaxies, and inspected
whether the correlations change along the radius. We find that the
isophotal shapes of the inner regions of our sample of galaxies are
statistically different from the isophotal shapes observed in the outer
regions. In the central regions we see patterns similar to those seen in
previous studies of nearby galaxies, with some galaxies showing 'boxy'
isophotes while others appear 'disky'. However, the pattern seen in the
inner region of each galaxy tends to change as the radius increases,
suggesting that while the inner parts of the galaxies are coherent and
presumably the result of specific dynamical processes, at larger radii the
shapes and orientations of the isophotes change and the behavior is not
well-defined. This may indicate effects
from the formation and evolution of each galaxy which are not yet fully
relaxed. However, for a full analysis of these data we need to know the
distances to the galaxies so that we can derive their luminosities, actual
sizes and other physical parameters. We hope to obtain fibre spectra for
many more galaxies to do this, preferably with one of the new Integral
Field Unit (IFU) systems that would avoid the effects of 'aperture bias'
that arise in single fibre spectra, and enable us to compare galaxies at
very different redshifts."

\section{Acknowledgments}
Two of the authors (LC and SKP) are grateful to ISRO for providing funds for this project under
their RESPOND scheme (project no. ISRO/RES/2/343/2007-08). A.M. was supported
by NSF grants AST-0407488 and AST-0909182. This paper
makes use of deep LFC data obtained at the California Institute
ofTechnology’s Palomar Observatories and we acknowledge the
role of S.G. Djorgovski,M. Bogosavljevic, and E. Glikman that
resulted in the data Facility: Hale Telescope.
We are thankful to Prof. Chanda Jog for giving valuble suggestions
through email communication regarding 
comparison of our work with results from N-body simulations.

We thank the anonymous referee for valuable comments. 

We thank the staff of the Australian Astronomical Observatory for
obtaining the AAT AAOmega spectra for us during Service Observing.

Funding for the Sloan Digital Sky Survey (SDSS) and SDSS-II has been 
provided by the Alfred P. Sloan Foundation, the Participating Institutions, 
the National Science Foundation, the U.S. Department of Energy, 
the National Aeronautics and Space Administration, the Japanese Monbukagakusho, 
and the Max Planck Society, and the Higher Education Funding Council for England. 
The SDSS Web site is http://www.sdss.org/.

The SDSS is managed by the Astrophysical Research Consortium (ARC) for the 
Participating Institutions. The Participating Institutions are the 
American Museum of Natural History, Astrophysical Institute Potsdam, 
University of Basel, University of Cambridge, Case Western Reserve University, 
The University of Chicago, Drexel University, Fermilab, the Institute for Advanced Study, 
the Japan Participation Group, The Johns Hopkins University, the 
Joint Institute for Nuclear Astrophysics, the Kavli Institute for Particle Astrophysics 
and Cosmology, the Korean Scientist Group, the Chinese Academy of Sciences (LAMOST), 
Los Alamos National Laboratory, the Max-Planck-Institute for Astronomy (MPIA), 
the Max-Planck-Institute for Astrophysics (MPA), New Mexico State University, 
Ohio State University, University of Pittsburgh, University of Portsmouth, 
Princeton University, the United States Naval Observatory, and the University of Washington.

\appendix

\section{Estimation of surface brightness limit with LFC image
as compared to the SDSS image}
\labsecn{calculation}
Following is the estimation of how far in the surface brightness we can go
with LFC images as compared to SDSS images assuming the poison statistics.

The number of source and background 
photons collected in t seconds using a telescope of diameter D is  
\begin{eqnarray}
N_{s} = n_{s}D^{2}t \nonumber \\
N_{b} = n_{b} D^{2} t
\labequn{No. of Photons}
\end{eqnarray}
where $n_{s}$ and $n_{b}$ are rates of photons received in unit area from 
source and the background. Signal-to-noise ratio (SNR) for an extended 
source which  spans a solid angle $\Omega$ can be given as
\begin{equation}
\frac{N_{s}}{\sigma_{s}} = \frac{n_{s}\left(\Omega D^{2}t 
\right)^\frac{1}{2}}{\sqrt{n_{s}+2n_{b}}}
\labequn{SNR}
\end{equation}
For $n_{s} \ll n_{b}$, the above equation becomes
\begin{equation}
\frac{N_{s}}{\sigma_{s}} = \frac{n_{s}\left(\Omega D^{2}t 
\right)^\frac{1}{2}}{\sqrt{2n_{b}}}
\labequn{SNR background limited}
\end{equation}
The SNR will increase by factor of $\sqrt{P}$, if the 
source intensity is obtained by averaging intensity in P number 
of pixels, 
\begin{equation}
SNR_{av} = \frac{N_{s}\sqrt{P}}{\sigma_{s}}
\labequn{SNR P}
\end{equation}
In ellipse fitting task to get the average intensity of an isophotes 
the intensity is averaged in elliptical annulus along the fitted ellipse.
The output table of \texttt{ellipse} gives parameters INTENS and INT\_ERR.
Where the  INTENS is the mean intensity along the fitted ellipse and the  INT\_ERR is the 
total error in the intensity. We can write   
\begin{equation}
\frac{INTENS}{INT\_ERR} = \frac{N_{s}\sqrt{P}}{\sigma_{s}}
\labequn{SNR ellipse}
\end{equation}
If the isophotes in two different images have same $\frac{INTENS}{INT\_ERR}$ then
\begin{equation}
\frac{n_{s1}\left(P_{1}\Omega_{1} D_{1}^{2}t_{1} 
\right)^\frac{1}{2}}{\sqrt{2n_{b1}}} = 
\frac{n_{s2}\left(P_{2}\Omega_{2} D_{2}^{2}t_{2} 
\right)^\frac{1}{2}}{\sqrt{2n_{b2}}}
\labequn{SNR two images}
\end{equation} 
The difference in surface brightness of isophotes ($\mu_{s1} - \mu_{s2}$) with 
same $\frac{INTENS}{INT\_ERR}$
in two different images can be given as 
\begin{equation}
\mu_{s1} - \mu_{s2} = \frac{1}{2}\left[ \mu_{b1}-\mu_{b2} 
- 2.5 log\left[  \left( \frac{D_{2}}{D_{1}}\right)^{2}\left( 
\frac{t_{2}}{t_{1}}\right) \left( \frac{\Omega_{2}}{\Omega_{1}}\right) 
\left(\frac{P_{2}}{P_{1}} \right) 
\right]  \right]
\labequn{surfacebrightness difference} 
\end{equation}
The ellipse is fitted till $\frac{INTENS}{INT\_ERR}$ drops to 3 to derive 
surface brightness profiles in the images of SDSS and LFC (see \fig{compare} as an example).
Putting Values for SDSS image: $D1$=2.5m, $t1$=54s, $P_{1}$=72 pixels, 
$\mu_{b1} = 20.3\magarcsecsqi $ and for 
LFC image: $D1$=5m, $t1$=9065s, $P_{1}$=184 pixels, 
$\mu_{b1} = 20.2 \magarcsecsqi$, 
in \equn{surfacebrightness difference} 
we expect to go minimum of $\sim 4 \magarcsecsqi$ deeper in the LFC 
image as compared to the SDSS image.

\newpage
\begin{deluxetable}{lllllllllllll}
\tablecolumns{13}
\tablewidth{0pc}
\rotate
\setlength{\tabcolsep}{0.02in}
\tabletypesize{\scriptsize}
\tablecaption{Basic parameters of the sample galaxies\labtablem{basic full}}
\tablehead{
\colhead{Our ID}    &  \colhead{SDSS ID}&
\colhead{RA} & \colhead{Dec} & 
\colhead{{\it u}} & \colhead{{\it g}} & \colhead{{\it r}} & \colhead{{\it i}} & \colhead{{\it z}}&
\colhead{$R_{50}$}& \colhead{sma} & \colhead{$\mu_{{\it i}}$} & \colhead{SNR} \\

\colhead{}    &  \colhead{}&
\colhead{}  & \colhead{} & 
\colhead{$\unit{mag}$}& \colhead{$\unit{mag}$} & \colhead{$\unit{mag}$}& \colhead{$\unit{mag}$}& \colhead{$\unit{mag}$} &
\colhead{$\unit{arcsec}$}& \colhead{$\unit{arcsec}$} & \colhead{$\magarcsecsqi$}& \colhead{}\\

\colhead{(1)}    &  \colhead{(2)}&
\colhead{(3)}  & \colhead{(4)} & 
\colhead{(5)}& \colhead{(6)} & \colhead{(7)}& \colhead{(8)}& \colhead{(9)} &
\colhead{(10)}& \colhead{(11)} & \colhead{(12)}& \colhead{(13)}

}
\startdata

LFC 462   & 588848899913548358 & 182.02545140 &  -0.00014042 & 21.50 $\pm$ 0.43 & 21.06 $\pm$ 0.13 & 20.55 $\pm$ 0.13 & 20.57 $\pm$ 0.21 & 19.96 $\pm$ 0.43 & 1.07 $\pm$ 0.16 & 3.81 & 26.36 $\pm$ 0.06 & 17.25 \\ 
LFC 525   & 587748929241612656 & 182.10068456 &  0.00579127 & 22.48 $\pm$ 1.58 & 20.76 $\pm$ 0.09 & 20.42 $\pm$ 0.10 & 20.13 $\pm$ 0.12 & 20.74 $\pm$ 0.95 & 1.10 $\pm$ 0.11 & 3.81 & 26.06 $\pm$ 0.04 & 26.07 \\ 
LFC 554   & 587748929241547219 & 181.99182097 &  0.00889758 & 22.09 $\pm$ 1.04 & 21.89 $\pm$ 0.24 & 21.11 $\pm$ 0.18 & 19.90 $\pm$ 0.09 & 19.67 $\pm$ 0.34 & 1.28 $\pm$ 0.07 & 3.81 & 26.67 $\pm$ 0.08 & 12.50 \\ 
LFC 558   & 587748929241547253 & 181.99959155 &  0.01032991 & 26.27 $\pm$ 2.14 & 21.73 $\pm$ 0.24 & 20.20 $\pm$ 0.09 & 19.44 $\pm$ 0.07 & 19.05 $\pm$ 0.21 & 1.01 $\pm$ 0.07 & 5.11 & 26.58 $\pm$ 0.10 & 10.06 \\ 
LFC 622   & 587748929241546923 & 182.03581581 &  0.01761600 & 20.79 $\pm$ 0.41 & 19.26 $\pm$ 0.03 & 18.16 $\pm$ 0.02 & 17.70 $\pm$ 0.02 & 17.33 $\pm$ 0.06 & 1.38 $\pm$ 0.02 & 10.11 & 27.21 $\pm$ 0.09 & 11.88 \\ 
LFC 633   & 587748929241547046 & 182.04195384 &  0.01802068 & 20.48 $\pm$ 0.30 & 20.09 $\pm$ 0.06 & 19.51 $\pm$ 0.05 & 19.11 $\pm$ 0.05 & 19.21 $\pm$ 0.28 & 1.26 $\pm$ 0.05 & 8.32 & 27.18 $\pm$ 0.08 & 13.77 \\ 
LFC 663   & 587748929241612845 & 182.17111251 &  0.01778466 & 21.46 $\pm$ 0.60 & 21.97 $\pm$ 0.26 & 20.89 $\pm$ 0.15 & 20.06 $\pm$ 0.10 & 19.19 $\pm$ 0.22 & 1.14 $\pm$ 0.08 & 3.81 & 25.90 $\pm$ 0.05 & 23.48 \\ 
LFC 677   & 587748929241547621 & 181.99102394 &  0.02035316 & 19.94 $\pm$ 0.20 & 22.14 $\pm$ 0.42 & 20.94 $\pm$ 0.16 & 20.98 $\pm$ 0.33 & 22.79 $\pm$ 4.34 & 0.80 $\pm$ 0.17 & 5.36 & 27.64 $\pm$ 0.19 & 5.11 \\
\enddata
\tablecomments{Column  (1) gives ID of the sample galaxies in SExtractor catalog generated
by us. Quantities from column (2) to column (10) are obtained from the 
 PhotoObjAll table of the SDSS DR7 photometric catalog \citep{aba09}.
Column (2) gives a unique SDSS identifier which is listed as objID in the PhotoObjAll table 
Columns (3) and (4) give J2000 right ascension and declination in degree. Columns (5) to (9)
give extinction corrected  Petrosian magnitudes in {\it u}, {\it g}, {\it r},{\it i} 
and {\it z} bands. Column (10) gives Petrosian half-light radius in a arcsec.
Columns (11) to (13) give length of semi-major axis (in arcsec), {\it i}-band surface brightness
(in $mag/arcsec^{-2}$) and signal to noise ratio respectively of the outermost isophote in which
ellipse can be fitted (see Section 4).
}
\end{deluxetable}

\begin{deluxetable}{llllll}
\tablecolumns{6}
\tablewidth{0pc}
\tabletypesize{\footnotesize}
\tablecaption{Derived parametrs for sample galaxies\labtablem{derived full}}
\tablehead{
\colhead{ID}    &  \colhead{B/T}&
\colhead{$\mean{\frac{a3}{a}}$} & \colhead{$\mean{\frac{a4}{a}}$} & 
\colhead{$\mean{\epsilon}$}& \colhead{$\delta PA $} \\

\colhead{(1)}    &  \colhead{(2)}&
\colhead{(3)} & \colhead{(4)} & 
\colhead{(5)}& \colhead{(6)}

}
\startdata
 LFC 462   & 0.89 & 	\nodata & \nodata & \nodata & \nodata \\ 
&& 3.52e-03 $\pm$ 2.25e-03 & -1.25e-03  $\pm$  2.16e-03 & 0.057  $\pm$  0.0058 & 60.37  $\pm$  3.02 \\ 
&& 7.01e-03 $\pm$ 4.73e-03 & 6.91e-03  $\pm$  3.09e-03 & 0.048  $\pm$  0.0063 & 1.87  $\pm$  3.67 \\ 
&& \nodata & \nodata & \nodata & \nodata \\ 
 
LFC 525   & 0.97 & 	\nodata & \nodata & \nodata & \nodata \\ 
&& 5.52e-03 $\pm$ 1.89e-03 & -1.09e-03  $\pm$  1.53e-03 & 0.070  $\pm$  0.0047 & 8.47  $\pm$  2.02 \\ 
&& -6.61e-03 $\pm$ 5.70e-03 & 1.21e-02  $\pm$  5.41e-03 & 0.023  $\pm$  0.0117 & 8.22  $\pm$  11.97 \\ 
&& \nodata & \nodata & \nodata & \nodata \\
\enddata
\tablecomments{ Column (1) gives our SExtractor catalog ID. Column (2) gives bulge-to-total luminosity
ratio (see Section 5). Columns (3) to (6) give average isophotal parameters 
$a_{3}/a$, $a_{4}/a$, ellipticity and isophotal twist (in $\unit{degree}$)  in four different regions 
for each galaxy in four rows. Average value of isophotal parameter in Region 1, Region 2, Region 3 and 
Region 4 is listed in first, second, third and fourth row 
respectively for each galaxy.     
}
\end{deluxetable}

\end{document}